\documentclass[twoside,twocolumn,9pt]{article}
\usepackage{extsizes}
\usepackage[super,sort&compress,comma]{natbib} 
\usepackage[version=3]{mhchem}
\usepackage[left=1.5cm, right=1.5cm, top=1.785cm, bottom=2.0cm]{geometry}
\usepackage{balance}
\usepackage{mathptmx}
\usepackage{sectsty}
\usepackage{graphicx} 
\usepackage{lastpage}
\usepackage[format=plain,justification=justified,singlelinecheck=false,font={stretch=1.125,small,sf},labelfont=bf,labelsep=space]{caption}
\usepackage{float}
\usepackage{fancyhdr}
\usepackage{fnpos}
\usepackage[english]{babel}
\addto{\captionsenglish}{%
  
}
\usepackage{array}
\usepackage{droidsans}
\usepackage{charter}
\usepackage[T1]{fontenc}
\usepackage[usenames,dvipsnames]{xcolor}
\usepackage{setspace}
\usepackage[compact]{titlesec}
\usepackage{hyperref}

\usepackage{epstopdf}

\definecolor{cream}{RGB}{222,217,201}

\begin{document}

\pagestyle{fancy}
\thispagestyle{plain}
\fancypagestyle{plain}{
\renewcommand{\headrulewidth}{0pt}
}

\makeFNbottom
\makeatletter
\renewcommand\LARGE{\@setfontsize\LARGE{15pt}{17}}
\renewcommand\Large{\@setfontsize\Large{12pt}{14}}
\renewcommand\large{\@setfontsize\large{10pt}{12}}
\renewcommand\footnotesize{\@setfontsize\footnotesize{7pt}{10}}
\makeatother

\renewcommand{\thefootnote}{\fnsymbol{footnote}}
\renewcommand\footnoterule{\vspace*{1pt}%
\color{cream}\hrule width 3.5in height 0.4pt \color{black}\vspace*{5pt}} 
\setcounter{secnumdepth}{5}

\makeatletter 
\renewcommand\@biblabel[1]{#1}            
\renewcommand\@makefntext[1]%
{\noindent\makebox[0pt][r]{\@thefnmark\,}#1}
\makeatother 
\renewcommand{\figurename}{\small{Fig.}~}
\sectionfont{\sffamily\Large}
\subsectionfont{\normalsize}
\subsubsectionfont{\bf}
\setstretch{1.125} 
\setlength{\skip\footins}{0.8cm}
\setlength{\footnotesep}{0.25cm}
\setlength{\jot}{10pt}
\titlespacing*{\section}{0pt}{4pt}{4pt}
\titlespacing*{\subsection}{0pt}{15pt}{1pt}

\fancyfoot{}
\fancyfoot[LO,RE]{\vspace{-7.1pt}\includegraphics[height=9pt]{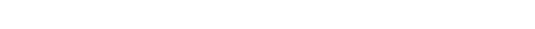}}
\fancyfoot[CO]{\vspace{-7.1pt}\hspace{13.2cm}\includegraphics{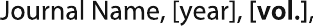}}
\fancyfoot[CE]{\vspace{-7.2pt}\hspace{-14.2cm}\includegraphics{head_foot/RF}}
\fancyfoot[RO]{\footnotesize{\sffamily{1--\pageref{LastPage} ~\textbar  \hspace{2pt}\thepage}}}
\fancyfoot[LE]{\footnotesize{\sffamily{\thepage~\textbar\hspace{3.45cm} 1--\pageref{LastPage}}}}
\fancyhead{}
\renewcommand{\headrulewidth}{0pt} 
\renewcommand{\footrulewidth}{0pt}
\setlength{\arrayrulewidth}{1pt}
\setlength{\columnsep}{6.5mm}
\setlength\bibsep{1pt}

\makeatletter 
\newlength{\figrulesep} 
\setlength{\figrulesep}{0.5\textfloatsep} 

\newcommand{\topfigrule}{\vspace*{-1pt}%
\noindent{\color{cream}\rule[-\figrulesep]{\columnwidth}{1.5pt}} }

\newcommand{\botfigrule}{\vspace*{-2pt}%
\noindent{\color{cream}\rule[\figrulesep]{\columnwidth}{1.5pt}} }

\newcommand{\dblfigrule}{\vspace*{-1pt}%
\noindent{\color{cream}\rule[-\figrulesep]{\textwidth}{1.5pt}} }

\makeatother

\twocolumn[
  \begin{@twocolumnfalse}
{\includegraphics[height=30pt]{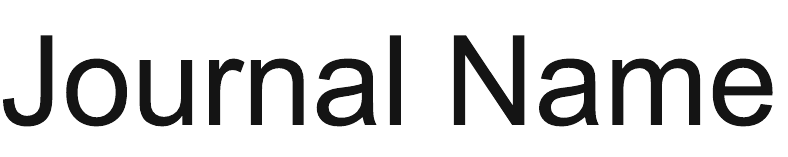}\hfill\raisebox{0pt}[0pt][0pt]{\includegraphics[height=55pt]{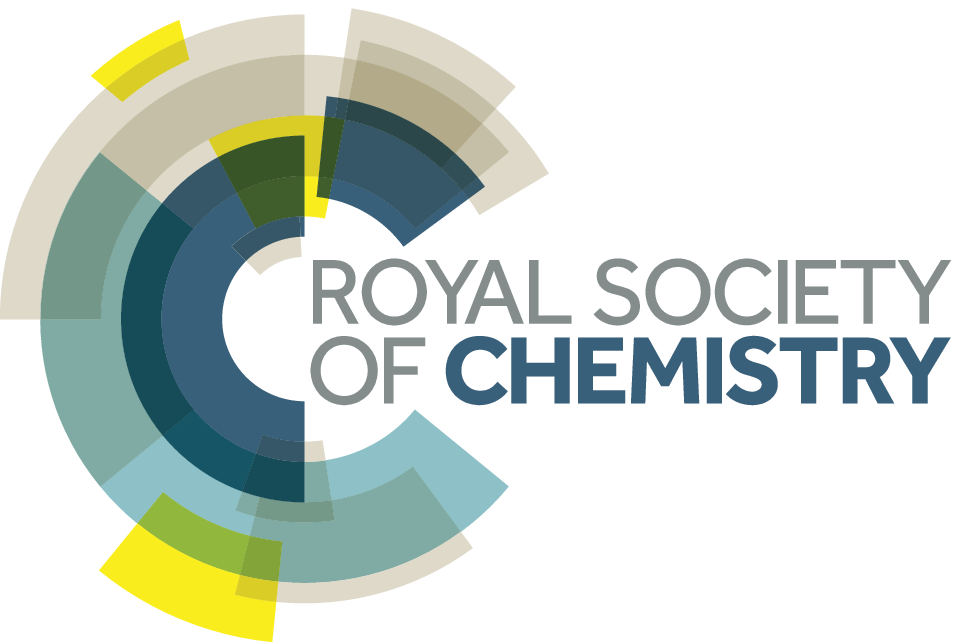}}\\[1ex]
\includegraphics[width=18.5cm]{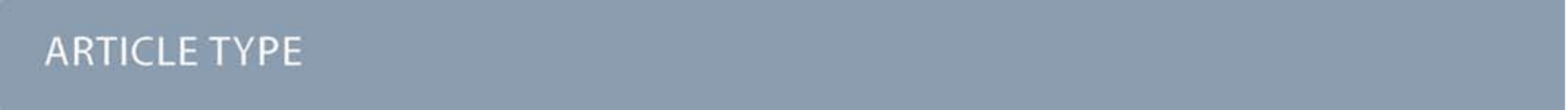}}\par
\vspace{1em}
\sffamily
\begin{tabular}{m{4.5cm} p{13.5cm} }

\includegraphics{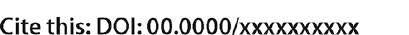} & \noindent\LARGE{\textbf{Manipulation of the electrical and memory properties of MoS$_2$ field-effect transistors by highly charged ion irradiation$^\dag$}} \\
\vspace{0.3cm} & \vspace{0.3cm} \\

& \noindent\large{Stephan Sleziona,$^{\ast}$\textit{$^{a}$} Aniello Pelella,\textit{$^{b}$} Enver Faella,\textit{$^{b}$} Osamah Kharsah,\textit{$^{a}$} Lucia Skopinski,\textit{$^{a}$} Andre Maas,\textit{$^{a}$} Yossarian Liebsch,\textit{$^{a}$} Antonio Di Bartolomeo,\textit{$^{b}$} and Marika Schleberger\textit{$^{a}$}} \\

\includegraphics{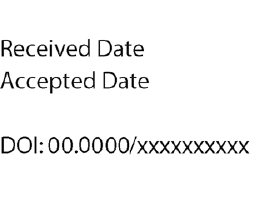} & \noindent\normalsize{Field-effect transistors based on molybdenum disulfide ($\mathrm{MoS_2}$) exhibit a hysteresis in their transfer characteristics, which can be utilized to realize 2D memory devices. This hysteresis has been attributed to charge trapping due to adsorbates, or defects either in the $\mathrm{MoS_2}$ lattice or in the underlying substrate. Wefabricated MoS$_2$ field-effect transistors on $\mathrm{SiO_2}$/Si substrates, irradiated these devices with $\mathrm{Xe^{30+}}$ ions at a kinetic energy of 180 keV to deliberately introduce defects and studied the resulting changes of their electrical and hysteretic properties. We find clear influences of the irradiation: While the charge carrier mobility decreases linearly with increasing ion fluence (up to only 20 \% of its initial value) the conductivity actually increases again after an initial drop of around two orders of magnitude, likely due to the occurence of hopping transport via localized states. We also find a significantly reduced $n$-doping ($\approx 10^{12}$ cm$^{-2}$) and a well-developed hysteresis after the irradiation. The hysteresis height increases with increasing ion fluence and enables us to characterize the irradiated $\mathrm{MoS_2}$ field-effect transistor as a memory device with remarkably longer relaxation times ($\approx$ minutes) compared to previous works.} \\

\end{tabular}

 \end{@twocolumnfalse} \vspace{0.6cm}

  ]

\renewcommand*\rmdefault{bch}\normalfont\upshape
\rmfamily
\section*{}
\vspace{-1cm}


\footnotetext{\textit{$^{a}$~Faculty of Physics and CENIDE, University of Duisburg-Essen, Lotharstraße 1, D-47057 Duisburg, Germany}}
\footnotetext{\textit{$^{b}$~Department of Physics “E. R. Caianiello”, University of Salerno, and CNR-SPIN, via Giovanni Paolo II, Fisciano 84084, Salerno, Italy}}




\section{Introduction}
Molybdenum disulfide ($\mathrm{MoS_2}$), a member of the family of the so-called transition metal dichalcogenides (TMDCs), is one of the most studied two-dimensional (2D) materials right after graphene. While in its bulk (3D) form it has an indirect bandgap of around 1.2 eV\cite{Kam.1982}, it develops a direct bandgap of 1.8-1.9 eV\cite{Mak.2010} when reduced to its covalently bonded S-Mo-S  monolayer structure. This bandgap allows the utilization of as typical building blocks for modern electronics like e.g., field-effect transistors (FETs) based on atomically thin 2D materials \cite{Radisavljevic.2011}. Because of that, it was quickly realized that monolayer $\mathrm{MoS_2}$ might be an excellent candidate for future electronic and opto-electronic applications, especially when large scale production techniques such as chemical vapor deposition (CVD) are used. The on-going reduction of device dimensions poses critical problems for traditional semiconductor devices e.g.~based on silicon, as the carrier mobility degrades rapidly for channel thicknesses reaching the scale of only a few nm \cite{Uchida.null,Gomez.2007,Schmidt.2009}, which is not the case for $\mathrm{MoS_2}$ and other 2D materials \cite{English.2016}. Indeed it was demonstrated that $\mathrm{MoS_2}$ FETs with a small gate length ($\leq$ 10 nm) and simultaneously reasonable mobility and high on-currents can be achieved \cite{English.2016b,Desai.2016}. Open challenges for 2D-TMDC FETs to date are Schottky barriers at the metal-TMDC interface \cite{Schulman.2018,Grillo.2021}, non-sufficient doping techniques \cite{He.2019} and structural defects either in the channel material or in the underlying oxide \cite{Knobloch.2022}. These structural defects can trap charges and act as scattering centres, modifying the electrical properties of the devices. One prominent consequence of these defects is the occurrence of a hysteresis in the transfer characteristics ($I_{\mathrm{DS}}(V_{\mathrm{GS}}$)) of a FET, which is commonly observed for $\mathrm{MoS_2}$ (and other TMDC) based devices \cite{Late.2012,Urban.2019,DiBartolomeo.2018,Shu.2016,Faella.2022}. The trapped charges influence the charge carriers in the 2D material channel and shift the transfer characteristics depending on the gate voltage sweep direction. Although most of the time the hysteresis should be prevented or eliminated for stable device performance, it can also be exploited to achieve atomically thin memory devices \cite{Bertolazzi.2013,He.2016,Kaushik.2017,Urban.2019}. 

Defects can be artificially and controllably introduced into 2D materials by particle irradiation, e.g. using electrons or ions as projectiles. While electron irradiation often leads to the creation of single vacancies \cite{Komsa.2013,Jin.2009}, ion irradiation can additionally cause more complex defects, depending on the ion type and energy \cite{Li.2017,Madau.2017,Ochedowski.2014,Kotakoski.2015}. These defects have been proposed or even utilized for a broad variety of applications, e.g. ultrafiltration \cite{CohenTanugi.2012,Madau.2017b}, DNA sequencing \cite{Postma.2010,Kozubek.2018} or catalysis \cite{Madau.2018}. By fine tuning the energy of the ions, irradiation can even be used to clean the surface of 2D materials from process residues stemming from transfer and lithography steps, without damaging the underlying 2D material too much \cite{Brennan.2021,Sleziona.2021}. Although there has been done some work with particle irradiation of 2D FETs \cite{Stanford.2016,Zion.2015,Li.2019,Bertolazzi.2017} these works focus mostly only on standard electrical performance like conductivity, mobility and their irradiation hardness \cite{Ernst.2016,Kumar.2018,Ochedowski.2013} and not on manipulating the hysteretic properties of the irradiated devices. 

With this work we want to close this gap and additionally pay specific attention to the manipulation of the hysteretic properties of 2D $\mathrm{MoS_2}$ FETs by ion irradiation. 
To this end we fabricate CVD-grown single layer $\mathrm{MoS_2}$ FETs on a $\mathrm{Si/SiO_2}$ substrate via photolithography and characterize their electrical properties by measuring their output and transfer characteristics. After this initial characterization, the devices are irradiated with highly charged ions (HCI) to deliberately introduce defects. We use HCIs as projectiles since their potential energy (e.g. their charge state) and kinetic energy can be tuned independently and by that control the defect creation in our devices. In this work we have chosen a relatively high kinetic energy so that we also expect defect creation in the substrate (see \cite{Skopinski.2023} for details). We show that the irradiation leads to distinct modifications of the electrical properties and especially causes a strongly reduced n-doping of the devices. Most importantly, we demonstrate that the irradiation leads to the opening of a hysteresis, most likely caused by additional defects in the underlying oxide. The height of the hysteresis scales with the introduced ion fluence, enabling the realisation of a memory device.

\section{Results and discussion}

\begin{figure*}[h]
\centering
\includegraphics{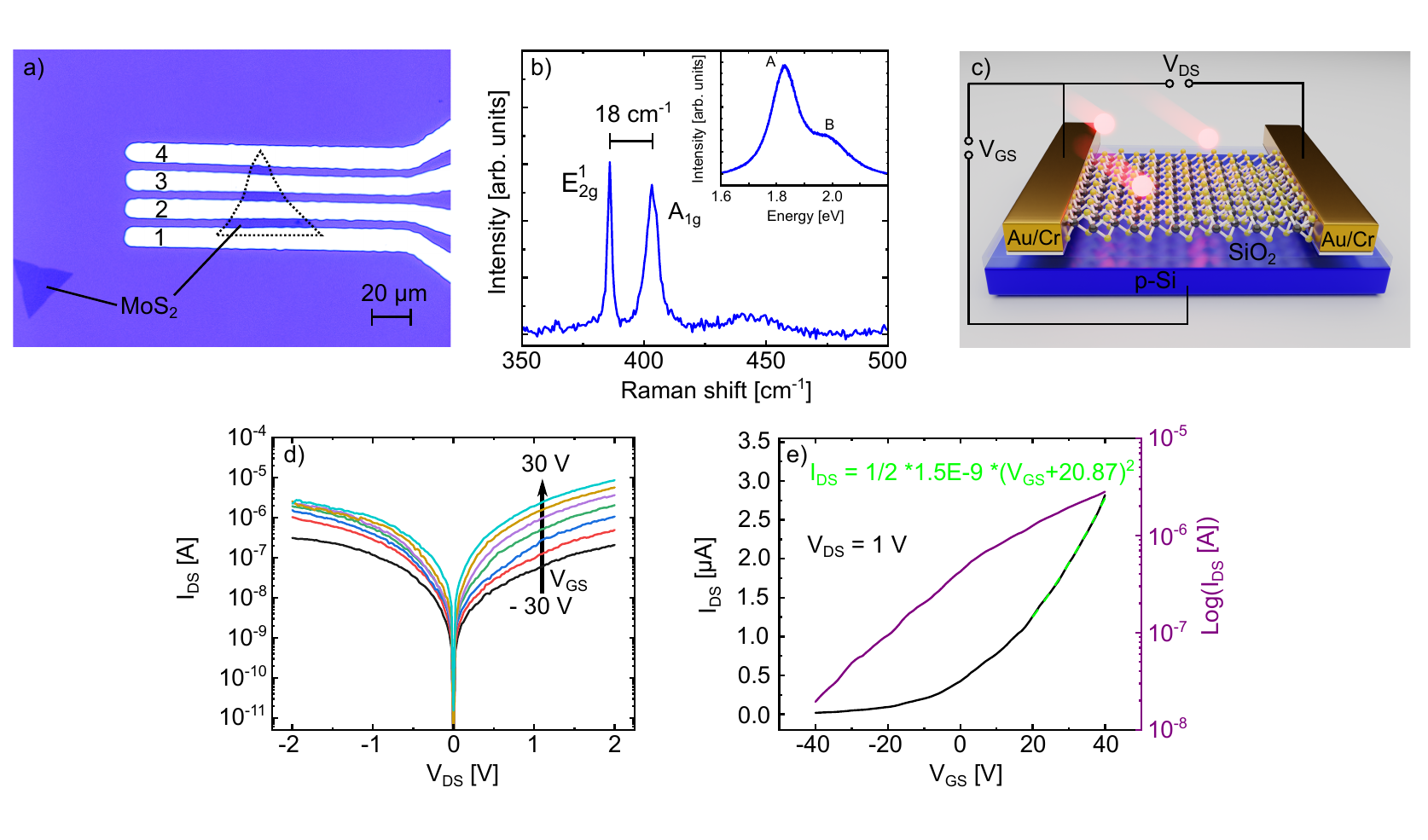}%
\caption{\label{Fig1} a) Optical microscopy image of monolayer $\mathrm{MoS_2}$ on $\mathrm{SiO_2}$ contacted with four Cr/Au leads. b) Raman spectrum of monolayer $\mathrm{MoS_2}$ with corresponding Photoluminescence spectrum as inset. c) Schematic of the device, measurement setup and general course of this work. d) Output and e) transfer characteristics of one of our devices before the irradiation.}
\end{figure*}

We begin by describing our devices and the general course of our work. In Fig. \ref{Fig1} a) an optical microscopy image of one of the $\mathrm{MoS_2}$-FETs used in this work is shown. The four metal contacts are labeled with numbers (\textbf{1}-\textbf{4}). For most devices, we employed a two-point measurement configuration where two contacts right next to each other were used as drain and source contact. For example, for the device shown in the image, the contacts \textbf{1} and \textbf{2} were used. Fig. \ref{Fig1} b) shows typical Raman and Photoluminescence (PL) data for our FETs after their fabrication. The two well-known $\mathrm{MoS_2}$ Raman modes $\mathrm{E^1_{2g}}$ and $\mathrm{A_{1g}}$ are present and the difference in their positions in the spectrum is $\mathrm{\approx 18}$ $\mathrm{cm^{-1}}$. The inset shows the PL spectrum measured at the same spatial position, displaying one strong peak at an energy of 1.83 eV attributed to the A exciton and a smaller peak at an energy of 1.98 eV attributed to the B exciton. Both observations clearly proof that our samples are indeed monolayers of $\mathrm{MoS_2}$ \cite{Lee.2010,Mak.2010}. Fig. \ref{Fig1} c) outlines the general course of our experiment: The FET structure is used for standard electrical characterization of the $\mathrm{MoS_2}$, in particular the output ($I_\mathrm{{DS}}$($V_\mathrm{{DS}}$)) and transfer ($I_\mathrm{{DS}}$($V_\mathrm{{GS}}$)) characteristics. After this initial characterization, the devices are irradiated with highly charged $\mathrm{Xe^{30+}}$ ions at a kinetic energy of 180 keV to deliberately introduce defects into the devices. Afterwards, the devices are again characterized to observe the influence of the introduced defects on their electrical behavior. We irradiated four different devices with four different fluences, namely 100 ions/$\mathrm{\mu}$m$^2$, 200 ions/$\mathrm{\mu}$m$^2$, 400 ions/$\mathrm{\mu}$m$^2$ and 1600 ions/$\mathrm{\mu}$m$^2$.

The output and transfer characteristics of one of our device before the irradiation is shown exemplary in Figure \ref{Fig1} d) and e), respectively. The output characteristics displays a slightly rectifying Schottky barrier between the channel and the contacts, which is a common observation for $\mathrm{MoS_2}$-FETs, since mid-gap Fermi level pinning arises from defects at various metal-TMDC interfaces caused e.g. by the processing conditions \cite{Kim.2017,Jung.2019,Guo.2015,Smyth.2016,Saidi.2014,DiBartolomeo.2018b}. The transfer characteristics in Figure \ref{Fig1} e) reveals the behavior of a normally-on n-type transistor with very strong n-doping. $\mathrm{MoS_2}$ is typically found to be n-doped, which is attributed to electron-donating sulfur vacancies \cite{Qiu.2013,Suh.2014,Tongay.2013,McDonnell.2014}. Consequently, the off-state of the transistor can not be reached in the applied gate voltage range, so the ratio between the minimum and maximum current is only $10^3$. We note that the devices in this work exhibit small differences in their overall electrical behavior, which is a typical observation for 2D devices in literature and can be explained by the contact resistance and Fermi-level pinning being delicately dependent on microscopic details in the contact formation at the metal-TMDC interface \cite{Liu.2018,Smyth.2016}. Nevertheless, all our devices have in common that they have a low Schottky barrier and exhibit strong n-doping. 

\begin{figure}
\includegraphics{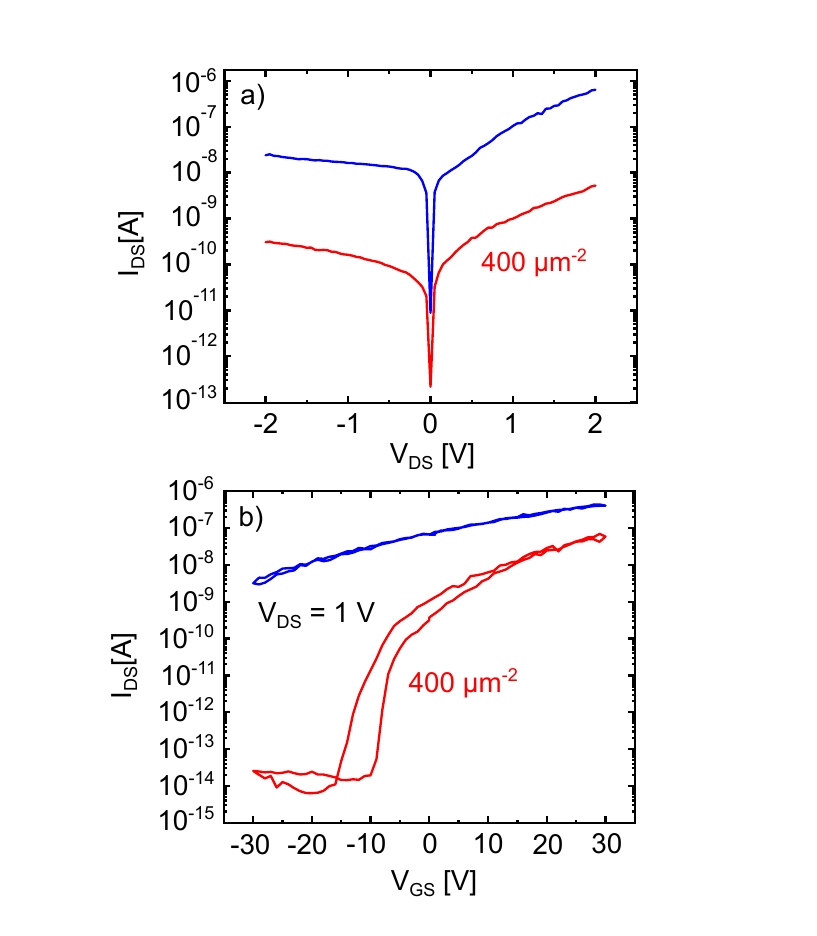}%
\caption{\label{Fig2} a) Output and b) transfer characteristics of a $\mathrm{MoS_2}$ FET before (blue) and after (red) irradiation with $\mathrm{Xe^{30+}}$ ions with a fluence of 400 ions/$\mathrm{\mu}$m$^2$.}%
\end{figure}

In Figure \ref{Fig2} a) and b) the output and transfer characteristics of the $\mathrm{MoS_2}$ device before (blue) and after irradiation (red) with a fluence of 400 ions/$\mathrm{\mu}$m$^2$ are shown. Let us discuss the output characteristic in Fig.~\ref{Fig2} (a) first. It displays a reduction of the current by around two orders of magnitude after the irradiation. Besides that, the Schottky type behavior is not modified. This finding is within our expectations: Particle irradiation of contacts can modify the metal-2D material interface in such devices and lead to reduced contact resistance and Schottky barriers \cite{Pelella.2020,Shahzad.2019}. In our work however, the kinetic energy of the ions is not high enough to penetrate through the metal contacts, as supported by SRIM calculations that demonstrate that all ion collision events occur only in the metal and not at the interface (see Figure S1 a)). This means that all the energy of the ions is deposited into the metal and not at the metal-TMDC interface from which follows, that the interface can not be modified by the irradiation. 

The strong reduction of $I_{\mathrm{DS}}$ indicates a significant increase in scattering centers. The irradiation of the $\mathrm{MoS_2}$ itself with electrons or ions with a moderate kinetic energy ($\approx$ keV) can lead to single or double vacancy defects in the TMDC lattice \cite{Lehtinen.2011,Lehtinen.2010,Zhou.2013}. In contrast, the defect creation mechanism by HCIs in 2D materials is still under discussion \cite{Schleberger.2018}. For free-standing $\mathrm{MoS_2}$, the formation of nm-sized holes after HCI irradiation was observed, where the size of the holes depends on the charge state of the ions \cite{Kozubek.2019}. Unfortunately to the best of our knowledge, there are no corresponding imaging experiments for substrate-supported $\mathrm{MoS_2}$ monolayers to date. However, recent time-of-flight mass spectrometry experiments show that the kinetic energy of the HCIs additionally has to be taken into account for substrate supported MoS$_2$ and that the HCI-MoS$_2$ interaction is also dependent on the type of the substrate \cite{Skopinski.2023}. 
Molecular dynamics simulations for substrate-supported $\mathrm{MoS_2}$ irradiated with cluster ions suggest, that there are also holes created in the $\mathrm{MoS_2}$ lattice, which have a bigger diameter than in the free-standing case \cite{Ghaderzadeh.2020}. Additionally, the substrate also gets modified, exhibiting, e.g., hillocks and thereby induce strain in the surrounding 2D material. Regardless of the specific defect type created by the irradiation (vacancies, holes, strained/chemically modified lattice), all of them will pose scattering centres for the charge carriers and thus reduce the conductivity of our device. \newline
We will now discuss the transfer characteristics in Figure \ref{Fig2} b) which shows striking differences between the measurement before (blue) and after (red) the irradiation. As discussed for Figure \ref{Fig1} e), the transfer characteristics before the irradiation displays a strong n-doping behavior. When sweeping the gate voltage between -30 V and +30 V and back, no hysteresis effect is observed. This might be caused by the strong n-doping of our devices, since the saturation region of the transfer characteristics usually does not show a significant hysteretic behavior when measured in high vacuum conditions \cite{Late.2012,DiBartolomeo.2018,Kaushik.2017}. After the irradiation, the most striking difference is the appearance of the off-state region and a hysteresis in the transfer characteristics. As mentioned before, this hysteresis is generally attributed to either defects in the $\mathrm{MoS_2}$ lattice, at the $\mathrm{MoS_2}$/oxide interface or in the oxide itself \cite{Kaushik.2017,DiBartolomeo.2018,Knobloch.2018}. As the occurrence of the hysteresis is clearly related to the irradiation it seems straightforward to claim that it is a result of additional defects introduced by the irradiation. The right-shift of the transfer after the forward gate voltage sweep, which leads to the clockwise hysteresis, is indicative of negative charge trapping. A further discussion of the properties of the hysteresis and which type of defect is likely the reason for its occurrence will be conducted later on. Additionally, the transfer curve shows, that the n-doping of the device is strongly reduced after the irradiation, since the threshold voltage ($V_{th}$) shifts towards positive gate voltages. It is now even possible to reach the off-state of the transistor in the applied gate voltage range and a high on-off ratio of nearly 6 orders of magnitude can be derived. 

This reduction of n-type doping may be attributed to several possible causes. Oxygen molecules capture electrons in $\mathrm{MoS_2}$ \cite{Nan.2014,Tongay.2013b,Wang.2022} and at the dangling bonds of the 2D material defect sites, the chemical and physical adsorption of molecules can be enhanced \cite{Wei.2014,Kong.2014,KC.2015}. Although the amount of adsorbed molecules should be reduced under vacuum conditions, ion irradiation can even lead to the formation of chemically adsorbed $\mathrm{MoO_3}$ at the defect edges, which would be very resistant to vacuum assisted desorption \cite{Madau.2018} and could also explain the observed reduction in n-doping. Since the HCIs will not only deposit their energy in the $\mathrm{MoS_2}$ monolayer, but also in the underlying $\mathrm{SiO_2}$ substrate, defects in the oxide could also play a role in p-doping the device. In fact, electron-trapping defect states for $\mathrm{MoS_2}$ on a $\mathrm{SiO_2}$ substrate have already been reported \cite{Knobloch.2018,Degraeve.2008,Illarionov.2017} and would lead to an effective p-doping of our devices by an additional gating effect. The trapped negative charges in the oxide influence the electric field generated by the applied gate voltage and thereby shifting the transfer curve towards positive gate voltages.

\begin{figure*}[h]
\centering
\includegraphics{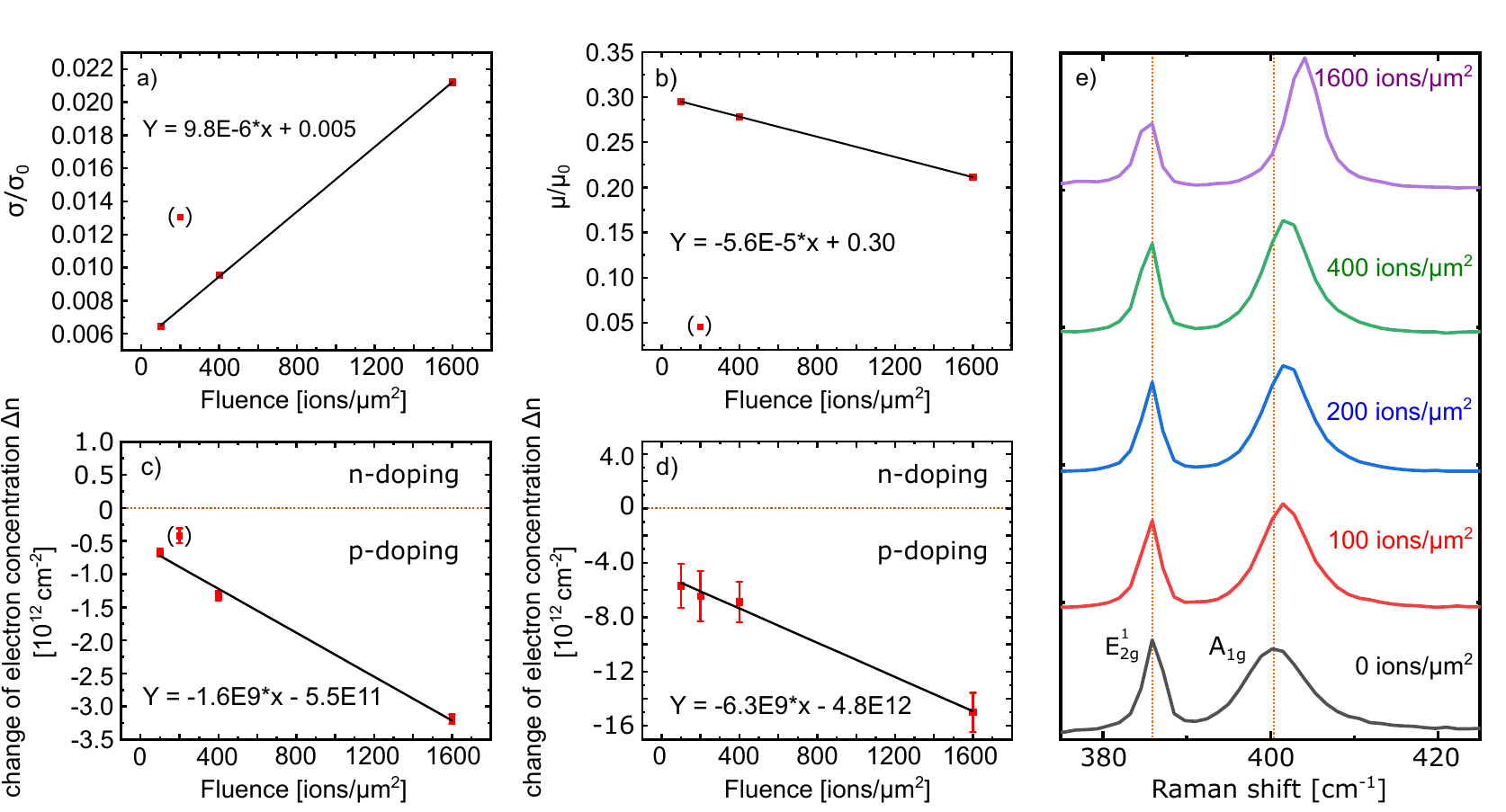}%
\caption{\label{Fig3} a) Remaining conductivity of the $\mathrm{MoS_2}$ FETs after the irradiation with different fluences normalized to the respective value of the device before the irradiation ($\sigma / \sigma_0$). b) Effective mobility of the MoS$\mathrm{_2}$ FETs after the irradiation with different fluences normalized to the respective value of the device before the irradiation ($\mu / \mu_0$). c) Change in charge carrier concentration of the MoS$\mathrm{_2}$ FETs for the different fluences calculated by the shift of $V_{\mathrm{th}}$. Negative values indicate a decrease of the electron density, meaning increased p-doping. d) Change in charge carrier concentration for the different fluences calculated from the Raman spectra in e) which are taken from a MoS$\mathrm{_2}$ sample at the different irradiation fluences.}
\end{figure*}

In the following, we will compare the results of the electrical characterization before and after the irradiation of our devices in dependence of the irradiation fluence. For this, we will address the conductivity, mobility, and charge carrier density, starting with the conductivity. In Figure \ref{Fig3} a) we display the remaining conductivity ($\sigma/\sigma_0$) using the output characteristics before and after irradiation. That is, we normalized the conductivity after irradiation to its value before irradiation to compare the different devices with each other. Interestingly, the devices show an increasing remaining conductivity with increasing ion fluence, after the initial drop in conductivity already discussed above (see Figure \ref{Fig2} a)). This peculiar behavior could be connected to an irradiation induced activation of an additional transport mechanism. For 2D TMDCs it has already been shown, that an increase in chalcogen vacancies or interface defects can lead to hopping transport via localized states and, as a consequence, lead to an increasing conductivity \cite{Qiu.2013,Huang.2020}. Note, that the FET irradiated with 200 ions/$\mathrm{\mu}$m$^2$ is the only exception to the otherwise linear behavior. For this device we used the contacts \textbf{1} and \textbf{3} as source and drain contact with one contact (2) in between on the $\mathrm{MoS_2}$ channel (see Figure \ref{Fig1} a)). By Fermi-level pinning this contact can modify the electrical behavior of the 2D material channel, which is mirrored in all electrical characterizations (Figure \ref{Fig3} a)-c)). We have therefore excluded this data point from our discussion. 

Next, we calculate the effective mobility by fitting the equation $I_{DS} = \frac{W}{L}\mu_{eff}Q_nV_{DS}$ to the transfer characteristics at high gate voltages ($>$ 20 V). With $Q_n = C_{ox} (V_{GS}-V_{th}$) and $C_{ox}$ = $1.21 \cdot 10^{-8}$ F/cm$^2$ we obtain values between 0.5-10 cm$^2$/Vs for the devices used in this work, which is in the range typically measured for such devices \cite{Novoselov.2005,Late.2012,Amani.2013}. After the irradiation, the mobility was examined again and then normalized to the value measured before the irradiation ($\mu / \mu_0$). The results of this analysis are shown in Figure \ref{Fig3} b) displaying a monotonous decrease of the mobility with increasing ion fluence. The defects introduced by the HCI irradiation either in the $\mathrm{MoS_2}$ or in the oxide can lead to increased Coulomb scattering for the charge carriers in our devices by charge trapping. This leads to shorter scattering times and therefore an overall reduced mobility, despite the enhanced remaining conductivity.

Lastly, we discuss the change in charge carrier density. We quantitatively evaluate the change in doping for our devices using the transfer characteristics of each device before and after irradiation. Because of the initially high n-doping of our devices we use $I_{DS} = \mu_v C_{ox} \frac{W}{L} (V_{GS} - V_{th})^2 V_{DS}$ to fit the transfer curve and extract the value for $V_{th}$ (see green-streaked line in Figure \ref{Fig1} e))\cite{DiBartolomeo.2018}. From this we calculated the change in charge carrier concentration with $\Delta n = C_{ox} \cdot \frac{\Delta V_{th}}{q}$. As can be seen  in Figure \ref{Fig3} c) , the irradiation leads to less n-doping (i.e. effective p-doping) in our devices in the order of $10^{12}$ cm$^{-2}$ and increases with increasing ion fluence up to 3.0 $\cdot 10^{-12}$ cm$^{-2}$ without any indication of a saturation behavior.

To further confirm this finding, we performed Raman spectroscopy of a CVD-grown $\mathrm{MoS_2}$ sample between the different irradiation steps (see Figure \ref{Fig3} e)). The qualitative behavior of the Raman spectra, a constant position for the $E^1_{2g}$ mode, while the $A_{1g}$ mode shifts to higher wavenumbers, points to decreasing n-doping \cite{Chakraborty.2012}, as it was also derived from the transfer characteristics. We also find a stronger reduction of the n-doping with increasing ion fluence. For a quantitative analysis we used the procedure from ref.~\cite{Pollmann.2020}. The result of this is shown in Figure \ref{Fig3} d). Obviously, both methods, electrical characterization and evaluation of Raman spectra, yield the same result, a linearly decreasing n-doping with increasing irradiation fluence, supporting our previous findings. The absolute values extracted from the Raman data are somewhat higher than those extracted from the transfer characteristics which can be explained by the fact that the Raman spectra were collected in ambient conditions, while the FETs were measured under high vacuum conditions. Therefore, the reduction of n-doping in MoS$_2$ by adsorbed oxygen plays a more important role in the Raman measurements than in case of the electrical measurements. 

We note, that the change in charge carrier density as derived by both methods is $\approx 10^{12} - 10^{13}$ cm$^{-2}$ and thus $1-2$ orders of magnitude higher than the irradiation fluence. As already discussed above, defect sites will facilitate p-doping. The defects we induce here are not point-like, but have a spatial extension on the order of nm. We therefore expect a high number of dangling bonds at the defect edges, which are prone to the adsorption or even bonding of gas molecules, explaining the high efficiency in terms of p-doping per ion. For the other possible cause of the observed doping effect, electron trapping defects in the SiO$_2$ substrate, there will also several defects per ion be created (see discussion below). This is also consistent with the the high efficiency in terms of doping per ion. Therefore, the observed doping effect can be explained satisfactory by both possibilities: either defects in the MoS$_2$ channel or in the underlying oxide. Since the measurements were performed under high vacuum conditions, defects in the oxide seem more likely.

\begin{figure*}[h]
\centering
\includegraphics{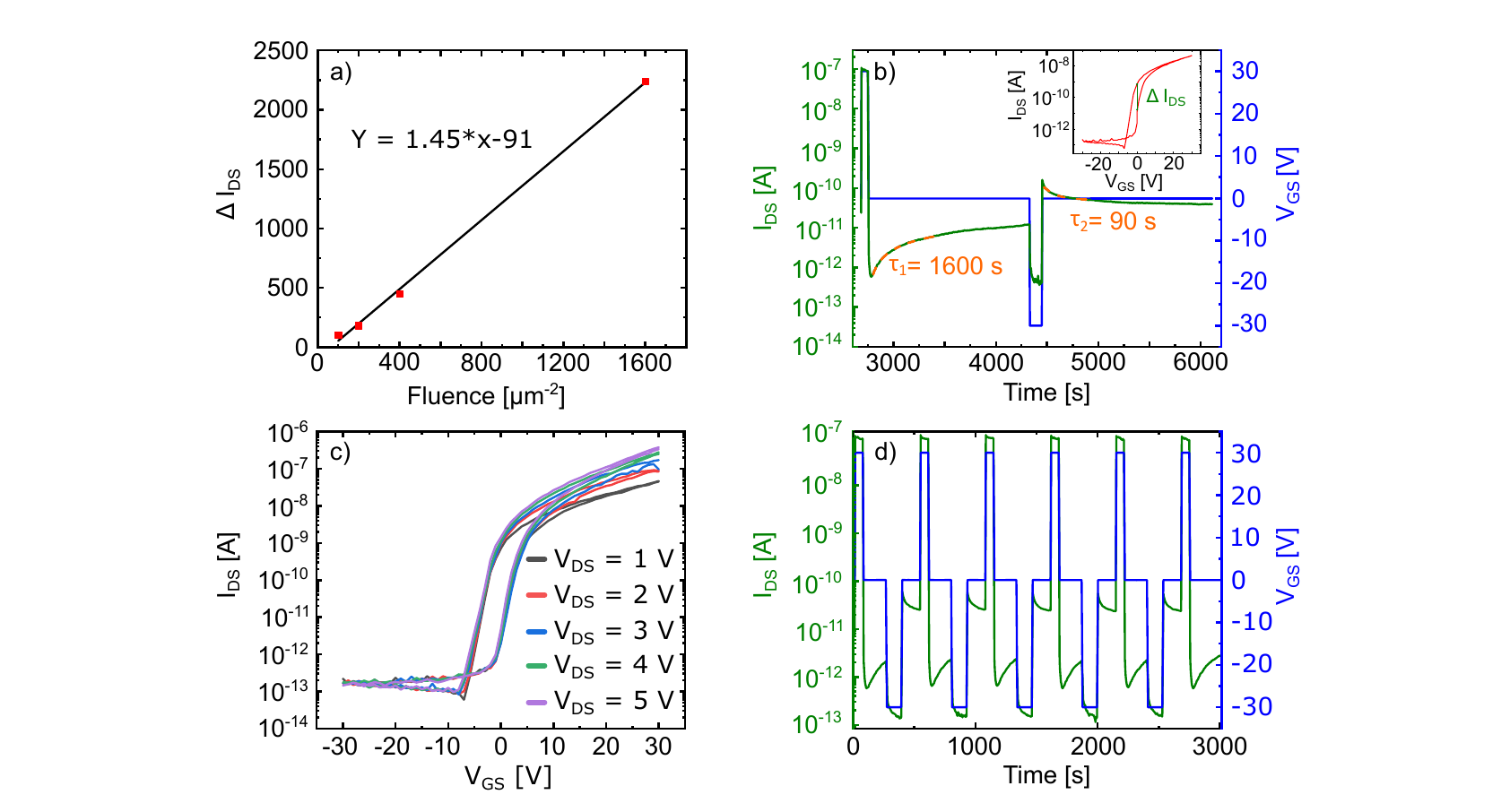}%
\caption{\label{Fig4} a) Hysteresis height (maximum $\Delta I_\mathrm{{DS}}$ at the same $V_\mathrm{{GS}}$, see also inset in b)) evaluated from the transfer characteristics of each device after the irradiation with different fluences. b) Transient behavior of the device irradiated with a fluence of 200 ions/$\mathrm{\mu}$m$^2$ for a single set-read-reset-read cycle. The dashed orange curves correspond to fits of exponential decays from which the trapping times $\tau_1$ and $\tau_2$ are evaluated. Inset shows the corresponding hysteresis curve with $\Delta I_\mathrm{{DS}}$ highlighted. c) Transfer characteristics of the device from b) for different values of ${V_\mathrm{DS}}$. d) Several set-read-reset-read cycles of the same device used in b).}
\end{figure*}

Finally, we want to address the manipulation of the hysteresis´ properties of the $\mathrm{MoS_2}$ devices by ion irradiation. As it was already shown in Figure \ref{Fig2} b) after the irradiation, a hysteresis can be observed in the devices transfer characteristics, which was absent before the irradiation. The origin of the hysteresis is generally attributed to defects, either in the $\mathrm{MoS_2}$ channel, the $\mathrm{MoS_2}$/oxide interface or in the oxide itself \cite{Late.2012,Guo.2015b,Choi.2015,Illarionov.2016,Knobloch.2018}. From an application point of view, such a device may be used as a non-volatile storage element. Favorable properties are two stable and clearly distinguishable memory states, the so-called memory window, a sufficiently high hysteresis to prevent unwanted switching and long time-constants when switched by erase/write voltage pulses.

For our analysis, we first evaluate the memory window, i.e., the height of the hysteresis (i.e. $\Delta I_{\mathrm{DS}}$ at the same $V_\mathrm{_{GS}}$) for the different irradiation fluences. The results found for our different devices are summarized in Figure \ref{Fig4} a). The hysteresis' height increases linearly with increasing ion fluence reaching up to around two orders of magnitude for the device irradiated with the highest fluence. We note, that even for the smallest irradiation fluence of only 100 ions/$\mathrm{\mu}$m$^2$ there is already a fully developed hysteresis observable even though it was nearly absent prior to the irradiation (see Figure S2). These results clearly demonstrate, that ion induced defects are responsible for the observed hysteresis.

To study the switching behavior of our device we applied $\pm$30 V gate pulses and recorded the transient behavior of the device irradiated with a fluence of 200 ions/$\mathrm{\mu}$m$^2$. The result is shown in Figure \ref{Fig4} b). In this device we can reach two distinct memory states at a gate voltage of $V_{GS}$ = 0 V with a current separation of around one order of magnitude. The current separation prevails and is stable for the entire observed pulse interval (around 30 min), which is comparable with the rentention times observed in few-layer MoS$_2$ charge-trapping memory devices \cite{Zhang.2015,Hou.2018}. The transients observed in Fig. \ref{Fig4} b) can be interpreted in terms of charge trapping/detrapping mechanisms. The time constants have been evaluated for the read ($\tau_1$) and the erase ($\tau_2$) configuration by fitting an exponential function $f(t)=c \cdot e^{-\frac{t}{\tau}} + A_0$ to the data. 

Compared to previous studies, the time constants $\mathrm{\tau_1}$ = 1600 s and $\mathrm{\tau_2}$ = 90 s, respectively, are rather long\cite{Urban.2019}. This finding points towards oxide defects playing a major role because charges trapped in deep oxide defects have considerably longer relaxation times than e.g. traps in the 2D material itself or at the 2D material/oxide interface \cite{Nagumo.06.12.201008.12.2010,Lee.2011,Kaushik.2017}. The time constant for the positive gate pulse ($\mathrm{\tau_1}$) is much longer than the time constant for negative gate pulse ($\mathrm{\tau_2}$). This is also another indicator of negatively charged oxide defects being the main contribution to the hysteresis in our work. These defects lie in the vicinity of the conduction band of MoS$_2$ \cite{Knobloch.2018} and would therefore be charged when applying positive gate voltages, but would not be charged for the negative gate pulse.  Additionally, we show in Figure \ref{Fig4} c) that the observed hysteresis is independent on the applied $V_\mathrm{{DS}}$ within the range of 1~V - 5~V. This is in contrast to recent observations for black phosphorous FETs, where the dependence of the hysteresis on $V_\mathrm{{DS}}$ was ascribed to defects in the 2D material channel itself \cite{Kumar.2023}.
Considering our irradiation conditions, significant defect generation in the substrate is to be expected. In Figure S1 b) we show a SRIM \cite{Ziegler.2010} calculation of our sample system with a monolayer $\mathrm{MoS_2}$ placed on top of a $\mathrm{SiO_2}$ layer irradiated with $\mathrm{Xe^+}$ ions with a kinetic energy of 180~keV. Most of the collisions caused by ion irradiation at this kinetic energy occur in the oxide, since the $\mathrm{MoS_2}$ channel is atomically thin.
We therefore conclude, that while the doping effect may be due to both, ion-induced defects in the MoS$_2$ and the substrate, the hysteresis observed in our devices is caused by negatively charged defects in the underlying oxide induced by the HCI irradiation. 
\newline
With Figure \ref{Fig4} d) we prove that the separation of the two memory states is stable for several memory cycles.

\section{Conclusion}
 We have investigated the manipulation of the electrical properties of $\mathrm{MoS_2}$ FETs by the irradiation with HCIs. While we found a decreasing mobility in the devices with increasing ion fluence, the conductivity after an initial drop actually increases with higher defect density suggesting that at hopping-like transport takes over with increasing defect density. This further proves, that the devices are rather resistant to ion irradiation, an important factor for the possible use in high radiation environments like e.g. space applications. Additionally, we have shown that HCI irradiation can be used for deliberate and controlled manipulation of the doping density  of $\mathrm{MoS_2}$ devices. In our case we found a strong decrease in n-doping. Most notably, the irradiation leads to a hysteresis in the transfer characteristics of the device which we successfully exploited for a non-volatile memory device with two stable memory states and a long retention time. We demonstrated that the memory window can be tuned by the irradiation fluence, opening up new possibilities to boost the performance of $\mathrm{MoS_2}$ based memory devices. We also believe, that this procedure can be applied to other similar 2D devices since the hysteresis probably originates from defects induced into the underlying oxide. In future experiments, it should be possible to precisely control the creation of defects and fine-tune the electrical properties through ion irradiation by independently controlling the potential and kinetic energy of the ions. This level of control would enable us to determine the specific locations where defects are created and thereby precisely manipulate the electrical properties of the irradiated devices.

\section{Experimental Procedure}
$\mathrm{MoS_2}$ flakes were grown via chemical vapor deposition (CVD) on a highly doped p-type Si substrate (resistivity 0.001-0.005 $\Omega$cm) covered by 285 nm thermal $\mathrm{SiO_2}$. At first a 1 \% sodium cholate solution was spin coated onto the substrate working as a seeding promoter. The growth was performed in a three-zone \linebreak (ThermConcept ROK 70/750/12-3z) tube furnace. By 10 min purging with 500~sccm Ar gas (99.9 \%) flow, the $\mathrm{O_2}$ content of the furnace was minimized. 40~mg of S powder (99.98 \% Sigma Aldrich) were placed in the upstream heating zone at 150 $^\circ$C. $\mathrm{MoO_3}$, used as the source for molybdenum, was obtained from a aqueous ammonium heptamolybdate (AHM) solution (ratio 1:1) initially annealed at 300 $^\circ$C for 24 min under ambient conditions and positioned in the next downstream zone at 750 $^\circ$C. During the whole process 500 sccm of Ar gas flows through the quartz tube. The growth process lasted 30 min and was followed by a rapid cooling. At a temperature of around 100 $^\circ$C the samples were retrieved from the CVD furnace. The resulting $\mathrm{MoS_2}$ flakes are mostly single layers with triangular shape.

For device fabrication the freshly grown samples were investigated via optical microscopy to select suitable flakes for photolithography processing. After the standard photolithography process 10 nm of Cr and 100 nm of Au were deposited by electron-beam (Cr) and thermal evaporation (Au) at a process pressure of $1 \cdot 10^{-5}$ mbar to electrically contact the $\mathrm{MoS_2}$ flakes.   

Electrical characterization of the devices was performed with a cryogenic probe station with pressure control and four metallic nanoprobes, which are connected to a Keithley 4200 SCS. The metallic sample plate was used to apply the backgate voltage to the Si substrate. All electrical measurements in this work are performed under a vacuum of $1 \cdot 10^{-4}$ mbar and the samples were left there for at least 12 hours before starting the measurements.

To irradiate the samples, highly charged xenon ions were generated in an electron beam ion source (EBIS) commercially available from Dreebit GmbH, Germany \cite{Zschornack.2008}. A kinetic energy of 180~keV (1.4~keV/amu) and an ion charge state of $q=30$ with a potential energy of 15.4~keV (0.1~keV/amu) was selected via a sector magnet and used for all experiments. Ion irradiation was performed under ultra-high vacuum conditions (pressure about $4 \cdot 10^{-9}$ mbar), and each sample was irradiated with a total fluence between 100 and 1600 ions/$\mu$m$^2$.

\section*{Author Contributions}
St.S., M.S. and A.d.B. are responsible for the conceptualization of the project and the experiments. St.S., A.P. and E.F. performed the investigation. St.S. analyzed and evaluated the data. L.S. performed the irradiation and helped with data interpretation. O.K., A.M. and Y.L. did the the device fabrication. M.S. and A.d.B. were responsible for funding acquisition and provided resources. St.S was responsible for writing the original draft. All authors contributed to the writing by reviewing and editing the manuscript.

\section*{Conflicts of interest}
There are no conflicts to declare.

\section*{Acknowledgments}
This work was funded by the Deutsche Forschungsgemeinschaft (DFG, German Research Foundation - project numbers INST-272132938, INST-429784087, 406129719 and 461605777 (IRTG 2038 2D MATURE) and the Bundesministerium für Bildung und Forschung (BMBF, project number 05K16PG1). We acknowledge support by the clean room staff of A. Lorke, especially G. Prinz.




\balance


\bibliography{Bibliography_irradiated_MoS2_FET} 

\providecommand*{\mcitethebibliography}{\thebibliography}
\csname @ifundefined\endcsname{endmcitethebibliography}
{\let\endmcitethebibliography\endthebibliography}{}
\begin{mcitethebibliography}{86}
\providecommand*{\natexlab}[1]{#1}
\providecommand*{\mciteSetBstSublistMode}[1]{}
\providecommand*{\mciteSetBstMaxWidthForm}[2]{}
\providecommand*{\mciteBstWouldAddEndPuncttrue}
  {\def\EndOfBibitem{\unskip.}}
\providecommand*{\mciteBstWouldAddEndPunctfalse}
  {\let\EndOfBibitem\relax}
\providecommand*{\mciteSetBstMidEndSepPunct}[3]{}
\providecommand*{\mciteSetBstSublistLabelBeginEnd}[3]{}
\providecommand*{\EndOfBibitem}{}
\mciteSetBstSublistMode{f}
\mciteSetBstMaxWidthForm{subitem}
{(\emph{\alph{mcitesubitemcount}})}
\mciteSetBstSublistLabelBeginEnd{\mcitemaxwidthsubitemform\space}
{\relax}{\relax}

\bibitem[Kam and Parkinson(1982)]{Kam.1982}
K.~K. Kam and B.~A. Parkinson, \emph{{The Journal of Physical Chemistry}},
  1982, \textbf{86}, 463--467\relax
\mciteBstWouldAddEndPuncttrue
\mciteSetBstMidEndSepPunct{\mcitedefaultmidpunct}
{\mcitedefaultendpunct}{\mcitedefaultseppunct}\relax
\EndOfBibitem
\bibitem[Mak \emph{et~al.}(2010)Mak, Lee, Hone, Shan, and Heinz]{Mak.2010}
K.~F. Mak, C.~Lee, J.~Hone, J.~Shan and T.~F. Heinz, \emph{{Physical review
  letters}}, 2010, \textbf{105}, 136805\relax
\mciteBstWouldAddEndPuncttrue
\mciteSetBstMidEndSepPunct{\mcitedefaultmidpunct}
{\mcitedefaultendpunct}{\mcitedefaultseppunct}\relax
\EndOfBibitem
\bibitem[Radisavljevic \emph{et~al.}(2011)Radisavljevic, Radenovic, Brivio,
  Giacometti, and Kis]{Radisavljevic.2011}
B.~Radisavljevic, A.~Radenovic, J.~Brivio, V.~Giacometti and A.~Kis,
  \emph{{Nature nanotechnology}}, 2011, \textbf{6}, 147--150\relax
\mciteBstWouldAddEndPuncttrue
\mciteSetBstMidEndSepPunct{\mcitedefaultmidpunct}
{\mcitedefaultendpunct}{\mcitedefaultseppunct}\relax
\EndOfBibitem
\bibitem[Uchida \emph{et~al.}(2002)Uchida, Watanabe, Kinoshita, Koga, Numata,
  and Takagi]{Uchida.null}
K.~Uchida, H.~Watanabe, A.~Kinoshita, J.~Koga, T.~Numata and S.~Takagi,
  {Digest. International Electron Devices Meeting}, 2002\relax
\mciteBstWouldAddEndPuncttrue
\mciteSetBstMidEndSepPunct{\mcitedefaultmidpunct}
{\mcitedefaultendpunct}{\mcitedefaultseppunct}\relax
\EndOfBibitem
\bibitem[Gomez \emph{et~al.}(2007)Gomez, berg, and Hoyt]{Gomez.2007}
L.~Gomez, I.~berg and J.~L. Hoyt, \emph{{IEEE Electron Device Letters}}, 2007,
  \textbf{28}, 285--287\relax
\mciteBstWouldAddEndPuncttrue
\mciteSetBstMidEndSepPunct{\mcitedefaultmidpunct}
{\mcitedefaultendpunct}{\mcitedefaultseppunct}\relax
\EndOfBibitem
\bibitem[Schmidt \emph{et~al.}(2009)Schmidt, Lemme, Gottlob, Driussi, Selmi,
  and Kurz]{Schmidt.2009}
M.~Schmidt, M.~C. Lemme, H.~Gottlob, F.~Driussi, L.~Selmi and H.~Kurz,
  \emph{{Solid-State Electronics}}, 2009, \textbf{53}, 1246--1251\relax
\mciteBstWouldAddEndPuncttrue
\mciteSetBstMidEndSepPunct{\mcitedefaultmidpunct}
{\mcitedefaultendpunct}{\mcitedefaultseppunct}\relax
\EndOfBibitem
\bibitem[English \emph{et~al.}(2016)English, Shine, Dorgan, Saraswat, and
  Pop]{English.2016}
C.~D. English, G.~Shine, V.~E. Dorgan, K.~C. Saraswat and E.~Pop, \emph{{Nano
  Letters}}, 2016, \textbf{16}, 3824--3830\relax
\mciteBstWouldAddEndPuncttrue
\mciteSetBstMidEndSepPunct{\mcitedefaultmidpunct}
{\mcitedefaultendpunct}{\mcitedefaultseppunct}\relax
\EndOfBibitem
\bibitem[English \emph{et~al.}(2016)English, Smithe, Xu, and
  Pop]{English.2016b}
C.~D. English, K.~K.~H. Smithe, R.~L. Xu and E.~Pop, {2016 International
  Electron Devices Meeting}, Piscataway, NJ, 2016, pp. 5.6.1--5.6.4\relax
\mciteBstWouldAddEndPuncttrue
\mciteSetBstMidEndSepPunct{\mcitedefaultmidpunct}
{\mcitedefaultendpunct}{\mcitedefaultseppunct}\relax
\EndOfBibitem
\bibitem[Desai \emph{et~al.}(2016)Desai, Madhvapathy, Sachid, Llinas, Wang,
  Ahn, Pitner, Kim, Bokor, Hu, Wong, and Javey]{Desai.2016}
S.~B. Desai, S.~R. Madhvapathy, A.~B. Sachid, J.~P. Llinas, Q.~Wang, G.~H. Ahn,
  G.~Pitner, M.~J. Kim, J.~Bokor, C.~Hu, H.-S.~P. Wong and A.~Javey,
  \emph{{Science}}, 2016, \textbf{354}, 99--102\relax
\mciteBstWouldAddEndPuncttrue
\mciteSetBstMidEndSepPunct{\mcitedefaultmidpunct}
{\mcitedefaultendpunct}{\mcitedefaultseppunct}\relax
\EndOfBibitem
\bibitem[Schulman \emph{et~al.}(2018)Schulman, Arnold, and Das]{Schulman.2018}
D.~S. Schulman, A.~J. Arnold and S.~Das, \emph{{Chemical Society reviews}},
  2018, \textbf{47}, 3037--3058\relax
\mciteBstWouldAddEndPuncttrue
\mciteSetBstMidEndSepPunct{\mcitedefaultmidpunct}
{\mcitedefaultendpunct}{\mcitedefaultseppunct}\relax
\EndOfBibitem
\bibitem[Grillo and {Di Bartolomeo}(2021)]{Grillo.2021}
A.~Grillo and A.~{Di Bartolomeo}, \emph{{Advanced Electronic Materials}}, 2021,
  \textbf{7}, 2000979\relax
\mciteBstWouldAddEndPuncttrue
\mciteSetBstMidEndSepPunct{\mcitedefaultmidpunct}
{\mcitedefaultendpunct}{\mcitedefaultseppunct}\relax
\EndOfBibitem
\bibitem[He \emph{et~al.}(2019)He, Liu, Tan, Zhai, Nam, and Zhang]{He.2019}
Q.~He, Y.~Liu, C.~Tan, W.~Zhai, G.-H. Nam and H.~Zhang, \emph{{ACS nano}},
  2019, \textbf{13}, 12294--12300\relax
\mciteBstWouldAddEndPuncttrue
\mciteSetBstMidEndSepPunct{\mcitedefaultmidpunct}
{\mcitedefaultendpunct}{\mcitedefaultseppunct}\relax
\EndOfBibitem
\bibitem[Knobloch \emph{et~al.}(2022)Knobloch, Selberherr, and
  Grasser]{Knobloch.2022}
T.~Knobloch, S.~Selberherr and T.~Grasser, \emph{{Nanomaterials}}, 2022,
  \textbf{12}, 3548\relax
\mciteBstWouldAddEndPuncttrue
\mciteSetBstMidEndSepPunct{\mcitedefaultmidpunct}
{\mcitedefaultendpunct}{\mcitedefaultseppunct}\relax
\EndOfBibitem
\bibitem[Late \emph{et~al.}(2012)Late, Liu, Matte, Dravid, and Rao]{Late.2012}
D.~J. Late, B.~Liu, H.~S. S.~R. Matte, V.~P. Dravid and C.~N.~R. Rao,
  \emph{{ACS nano}}, 2012, \textbf{6}, 5635--5641\relax
\mciteBstWouldAddEndPuncttrue
\mciteSetBstMidEndSepPunct{\mcitedefaultmidpunct}
{\mcitedefaultendpunct}{\mcitedefaultseppunct}\relax
\EndOfBibitem
\bibitem[Urban \emph{et~al.}(2019)Urban, Giubileo, Grillo, Iemmo, Luongo,
  Passacantando, Foller, Madau{\ss}, Pollmann, Geller, Oing, Schleberger, and
  {Di Bartolomeo}]{Urban.2019}
F.~Urban, F.~Giubileo, A.~Grillo, L.~Iemmo, G.~Luongo, M.~Passacantando,
  T.~Foller, L.~Madau{\ss}, E.~Pollmann, M.~P. Geller, D.~Oing, M.~Schleberger
  and A.~{Di Bartolomeo}, \emph{{2D Materials}}, 2019, \textbf{6}, 045049\relax
\mciteBstWouldAddEndPuncttrue
\mciteSetBstMidEndSepPunct{\mcitedefaultmidpunct}
{\mcitedefaultendpunct}{\mcitedefaultseppunct}\relax
\EndOfBibitem
\bibitem[{Di Bartolomeo} \emph{et~al.}(2018){Di Bartolomeo}, Genovese,
  Giubileo, Iemmo, Luongo, Foller, and Schleberger]{DiBartolomeo.2018}
A.~{Di Bartolomeo}, L.~Genovese, F.~Giubileo, L.~Iemmo, G.~Luongo, T.~Foller
  and M.~Schleberger, \emph{{2D Materials}}, 2018, \textbf{5}, 015014\relax
\mciteBstWouldAddEndPuncttrue
\mciteSetBstMidEndSepPunct{\mcitedefaultmidpunct}
{\mcitedefaultendpunct}{\mcitedefaultseppunct}\relax
\EndOfBibitem
\bibitem[Shu \emph{et~al.}(2016)Shu, Wu, Guo, Liu, Wei, and Chen]{Shu.2016}
J.~Shu, G.~Wu, Y.~Guo, B.~Liu, X.~Wei and Q.~Chen, \emph{{Nanoscale}}, 2016,
  \textbf{8}, 3049--3056\relax
\mciteBstWouldAddEndPuncttrue
\mciteSetBstMidEndSepPunct{\mcitedefaultmidpunct}
{\mcitedefaultendpunct}{\mcitedefaultseppunct}\relax
\EndOfBibitem
\bibitem[Faella \emph{et~al.}(2022)Faella, Intonti, Viscardi, Giubileo, Kumar,
  Lam, Anastasiou, Craciun, Russo, and {Di Bartolomeo}]{Faella.2022}
E.~Faella, K.~Intonti, L.~Viscardi, F.~Giubileo, A.~Kumar, H.~T. Lam,
  K.~Anastasiou, M.~F. Craciun, S.~Russo and A.~{Di Bartolomeo},
  \emph{{Nanomaterials}}, 2022, \textbf{12}, 1886\relax
\mciteBstWouldAddEndPuncttrue
\mciteSetBstMidEndSepPunct{\mcitedefaultmidpunct}
{\mcitedefaultendpunct}{\mcitedefaultseppunct}\relax
\EndOfBibitem
\bibitem[Bertolazzi \emph{et~al.}(2013)Bertolazzi, Krasnozhon, and
  Kis]{Bertolazzi.2013}
S.~Bertolazzi, D.~Krasnozhon and A.~Kis, \emph{{ACS nano}}, 2013, \textbf{7},
  3246--3252\relax
\mciteBstWouldAddEndPuncttrue
\mciteSetBstMidEndSepPunct{\mcitedefaultmidpunct}
{\mcitedefaultendpunct}{\mcitedefaultseppunct}\relax
\EndOfBibitem
\bibitem[He \emph{et~al.}(2016)He, Ramamoorthy, Kwan, Lee, Nathawat,
  Somphonsane, Matsunaga, Higuchi, Yamanaka, Aoki, Gong, Zhang, Vajtai, Ajayan,
  and Bird]{He.2016}
G.~He, H.~Ramamoorthy, C.-P. Kwan, Y.-H. Lee, J.~Nathawat, R.~Somphonsane,
  M.~Matsunaga, A.~Higuchi, T.~Yamanaka, N.~Aoki, Y.~Gong, X.~Zhang, R.~Vajtai,
  P.~M. Ajayan and J.~P. Bird, \emph{{Nano Letters}}, 2016, \textbf{16},
  6445--6451\relax
\mciteBstWouldAddEndPuncttrue
\mciteSetBstMidEndSepPunct{\mcitedefaultmidpunct}
{\mcitedefaultendpunct}{\mcitedefaultseppunct}\relax
\EndOfBibitem
\bibitem[Kaushik \emph{et~al.}(2017)Kaushik, Mackenzie, Thakar, Goyal,
  Mukherjee, Boggild, Petersen, and Lodha]{Kaushik.2017}
N.~Kaushik, D.~M.~A. Mackenzie, K.~Thakar, N.~Goyal, B.~Mukherjee, P.~Boggild,
  D.~H. Petersen and S.~Lodha, \emph{{npj 2D Materials and Applications}},
  2017, \textbf{1}, 34\relax
\mciteBstWouldAddEndPuncttrue
\mciteSetBstMidEndSepPunct{\mcitedefaultmidpunct}
{\mcitedefaultendpunct}{\mcitedefaultseppunct}\relax
\EndOfBibitem
\bibitem[Komsa \emph{et~al.}(2013)Komsa, Kurasch, Lehtinen, Kaiser, and
  Krasheninnikov]{Komsa.2013}
H.-P. Komsa, S.~Kurasch, O.~Lehtinen, U.~Kaiser and A.~V. Krasheninnikov,
  \emph{{Physical Review B}}, 2013, \textbf{88}, 035301\relax
\mciteBstWouldAddEndPuncttrue
\mciteSetBstMidEndSepPunct{\mcitedefaultmidpunct}
{\mcitedefaultendpunct}{\mcitedefaultseppunct}\relax
\EndOfBibitem
\bibitem[Jin \emph{et~al.}(2009)Jin, Lin, Suenaga, and Iijima]{Jin.2009}
C.~Jin, F.~Lin, K.~Suenaga and S.~Iijima, \emph{{Physical review letters}},
  2009, \textbf{102}, 195505\relax
\mciteBstWouldAddEndPuncttrue
\mciteSetBstMidEndSepPunct{\mcitedefaultmidpunct}
{\mcitedefaultendpunct}{\mcitedefaultseppunct}\relax
\EndOfBibitem
\bibitem[Li and Chen(2017)]{Li.2017}
Z.~Li and F.~Chen, \emph{{Applied Physics Reviews}}, 2017, \textbf{4},
  011103\relax
\mciteBstWouldAddEndPuncttrue
\mciteSetBstMidEndSepPunct{\mcitedefaultmidpunct}
{\mcitedefaultendpunct}{\mcitedefaultseppunct}\relax
\EndOfBibitem
\bibitem[Madau{\ss} \emph{et~al.}(2017)Madau{\ss}, Ochedowski, Lebius,
  Ban-d'Etat, Naylor, Johnson, Kotakoski, and Schleberger]{Madau.2017}
L.~Madau{\ss}, O.~Ochedowski, H.~Lebius, B.~Ban-d'Etat, C.~H. Naylor, A.~T.~C.
  Johnson, J.~Kotakoski and M.~Schleberger, \emph{{2D Materials}}, 2017,
  \textbf{4}, 015034\relax
\mciteBstWouldAddEndPuncttrue
\mciteSetBstMidEndSepPunct{\mcitedefaultmidpunct}
{\mcitedefaultendpunct}{\mcitedefaultseppunct}\relax
\EndOfBibitem
\bibitem[Ochedowski \emph{et~al.}(2014)Ochedowski, Bukowska, {Freire Soler},
  Br{\"o}kers, Ban-d'Etat, Lebius, and Schleberger]{Ochedowski.2014}
O.~Ochedowski, H.~Bukowska, V.~M. {Freire Soler}, L.~Br{\"o}kers,
  B.~Ban-d'Etat, H.~Lebius and M.~Schleberger, \emph{{Nuclear Instruments and
  Methods in Physics Research Section B: Beam Interactions with Materials and
  Atoms}}, 2014, \textbf{340}, 39--43\relax
\mciteBstWouldAddEndPuncttrue
\mciteSetBstMidEndSepPunct{\mcitedefaultmidpunct}
{\mcitedefaultendpunct}{\mcitedefaultseppunct}\relax
\EndOfBibitem
\bibitem[Kotakoski \emph{et~al.}(2015)Kotakoski, Brand, Lilach, Cheshnovsky,
  Mangler, Arndt, and Meyer]{Kotakoski.2015}
J.~Kotakoski, C.~Brand, Y.~Lilach, O.~Cheshnovsky, C.~Mangler, M.~Arndt and
  J.~C. Meyer, \emph{{Nano Letters}}, 2015, \textbf{15}, 5944--5949\relax
\mciteBstWouldAddEndPuncttrue
\mciteSetBstMidEndSepPunct{\mcitedefaultmidpunct}
{\mcitedefaultendpunct}{\mcitedefaultseppunct}\relax
\EndOfBibitem
\bibitem[Cohen-Tanugi and Grossman(2012)]{CohenTanugi.2012}
D.~Cohen-Tanugi and J.~C. Grossman, \emph{{Nano Letters}}, 2012, \textbf{12},
  3602--3608\relax
\mciteBstWouldAddEndPuncttrue
\mciteSetBstMidEndSepPunct{\mcitedefaultmidpunct}
{\mcitedefaultendpunct}{\mcitedefaultseppunct}\relax
\EndOfBibitem
\bibitem[Madau{\ss} \emph{et~al.}(2017)Madau{\ss}, Schumacher, Ghosh,
  Ochedowski, Meyer, Lebius, Ban-d'Etat, Toimil-Molares, Trautmann, Lammertink,
  Ulbricht, and Schleberger]{Madau.2017b}
L.~Madau{\ss}, J.~Schumacher, M.~Ghosh, O.~Ochedowski, J.~Meyer, H.~Lebius,
  B.~Ban-d'Etat, M.~E. Toimil-Molares, C.~Trautmann, R.~G.~H. Lammertink,
  M.~Ulbricht and M.~Schleberger, \emph{{Nanoscale}}, 2017, \textbf{9},
  10487--10493\relax
\mciteBstWouldAddEndPuncttrue
\mciteSetBstMidEndSepPunct{\mcitedefaultmidpunct}
{\mcitedefaultendpunct}{\mcitedefaultseppunct}\relax
\EndOfBibitem
\bibitem[Postma(2010)]{Postma.2010}
H.~W.~C. Postma, \emph{{Nano Letters}}, 2010, \textbf{10}, 420--425\relax
\mciteBstWouldAddEndPuncttrue
\mciteSetBstMidEndSepPunct{\mcitedefaultmidpunct}
{\mcitedefaultendpunct}{\mcitedefaultseppunct}\relax
\EndOfBibitem
\bibitem[Kozubek \emph{et~al.}(2018)Kozubek, Ernst, Herbig, Michely, and
  Schleberger]{Kozubek.2018}
R.~Kozubek, P.~Ernst, C.~Herbig, T.~Michely and M.~Schleberger, \emph{{ACS
  Applied Nano Materials}}, 2018, \textbf{1}, 3765--3773\relax
\mciteBstWouldAddEndPuncttrue
\mciteSetBstMidEndSepPunct{\mcitedefaultmidpunct}
{\mcitedefaultendpunct}{\mcitedefaultseppunct}\relax
\EndOfBibitem
\bibitem[Madau{\ss} \emph{et~al.}(2018)Madau{\ss}, Zegkinoglou, {V{\'a}zquez
  Mui{\~n}os}, Choi, Kunze, Zhao, Naylor, Ernst, Pollmann, Ochedowski, Lebius,
  Benyagoub, Ban-d'Etat, Johnson, Djurabekova, {Roldan Cuenya}, and
  Schleberger]{Madau.2018}
L.~Madau{\ss}, I.~Zegkinoglou, H.~{V{\'a}zquez Mui{\~n}os}, Y.-W. Choi,
  S.~Kunze, M.-Q. Zhao, C.~H. Naylor, P.~Ernst, E.~Pollmann, O.~Ochedowski,
  H.~Lebius, A.~Benyagoub, B.~Ban-d'Etat, A.~T.~C. Johnson, F.~Djurabekova,
  B.~{Roldan Cuenya} and M.~Schleberger, \emph{{Nanoscale}}, 2018, \textbf{10},
  22908--22916\relax
\mciteBstWouldAddEndPuncttrue
\mciteSetBstMidEndSepPunct{\mcitedefaultmidpunct}
{\mcitedefaultendpunct}{\mcitedefaultseppunct}\relax
\EndOfBibitem
\bibitem[Brennan \emph{et~al.}(2021)Brennan, Centeno, Zurutuza, Mack, Paton,
  and Pollard]{Brennan.2021}
B.~Brennan, A.~Centeno, A.~Zurutuza, P.~Mack, K.~R. Paton and A.~J. Pollard,
  \emph{{ACS Applied Nano Materials}}, 2021, \textbf{4}, 5187--5197\relax
\mciteBstWouldAddEndPuncttrue
\mciteSetBstMidEndSepPunct{\mcitedefaultmidpunct}
{\mcitedefaultendpunct}{\mcitedefaultseppunct}\relax
\EndOfBibitem
\bibitem[Sleziona \emph{et~al.}(2021)Sleziona, Rauls, Heckhoff, Christen,
  Pollmann, Madau{\ss}, Franzka, Lorke, Wende, and Schleberger]{Sleziona.2021}
S.~Sleziona, S.~Rauls, T.~Heckhoff, L.~Christen, E.~Pollmann, L.~Madau{\ss},
  S.~Franzka, A.~Lorke, H.~Wende and M.~Schleberger, \emph{{Nanotechnology}},
  2021, \textbf{32}, 205702\relax
\mciteBstWouldAddEndPuncttrue
\mciteSetBstMidEndSepPunct{\mcitedefaultmidpunct}
{\mcitedefaultendpunct}{\mcitedefaultseppunct}\relax
\EndOfBibitem
\bibitem[Stanford \emph{et~al.}(2016)Stanford, Pudasaini, Belianinov, Cross,
  Noh, Koehler, Mandrus, Duscher, Rondinone, Ivanov, Ward, and
  Rack]{Stanford.2016}
M.~G. Stanford, P.~R. Pudasaini, A.~Belianinov, N.~Cross, J.~H. Noh, M.~R.
  Koehler, D.~G. Mandrus, G.~Duscher, A.~J. Rondinone, I.~N. Ivanov, T.~Z. Ward
  and P.~D. Rack, \emph{{Scientific reports}}, 2016, \textbf{6}, 27276\relax
\mciteBstWouldAddEndPuncttrue
\mciteSetBstMidEndSepPunct{\mcitedefaultmidpunct}
{\mcitedefaultendpunct}{\mcitedefaultseppunct}\relax
\EndOfBibitem
\bibitem[Zion \emph{et~al.}(2015)Zion, Haran, Butenko, Wolfson, Kaganovskii,
  Havdala, Sharoni, Naveh, Richter, Kaveh, Kogan, and Shlimak]{Zion.2015}
E.~Zion, A.~Haran, A.~Butenko, L.~Wolfson, Y.~Kaganovskii, T.~Havdala,
  A.~Sharoni, D.~Naveh, V.~Richter, M.~Kaveh, E.~Kogan and I.~Shlimak,
  \emph{{Graphene}}, 2015, \textbf{04}, 45--53\relax
\mciteBstWouldAddEndPuncttrue
\mciteSetBstMidEndSepPunct{\mcitedefaultmidpunct}
{\mcitedefaultendpunct}{\mcitedefaultseppunct}\relax
\EndOfBibitem
\bibitem[Li \emph{et~al.}(2019)Li, Liu, Zhang, Qi, Wei, Zhou, Wang, Ma, Tsai,
  Dong, and Huo]{Li.2019}
H.~Li, C.~Liu, Y.~Zhang, C.~Qi, Y.~Wei, J.~Zhou, T.~Wang, G.~Ma, H.-S. Tsai,
  S.~Dong and M.~Huo, \emph{{Nanotechnology}}, 2019, \textbf{30}, 485201\relax
\mciteBstWouldAddEndPuncttrue
\mciteSetBstMidEndSepPunct{\mcitedefaultmidpunct}
{\mcitedefaultendpunct}{\mcitedefaultseppunct}\relax
\EndOfBibitem
\bibitem[Bertolazzi \emph{et~al.}(2017)Bertolazzi, Bonacchi, Nan, Pershin,
  Beljonne, and Samor{\`i}]{Bertolazzi.2017}
S.~Bertolazzi, S.~Bonacchi, G.~Nan, A.~Pershin, D.~Beljonne and P.~Samor{\`i},
  \emph{{Advanced Materials}}, 2017, \textbf{29}, 1606760\relax
\mciteBstWouldAddEndPuncttrue
\mciteSetBstMidEndSepPunct{\mcitedefaultmidpunct}
{\mcitedefaultendpunct}{\mcitedefaultseppunct}\relax
\EndOfBibitem
\bibitem[Ernst \emph{et~al.}(2016)Ernst, Kozubek, Madau{\ss}, Sonntag, Lorke,
  and Schleberger]{Ernst.2016}
P.~Ernst, R.~Kozubek, L.~Madau{\ss}, J.~Sonntag, A.~Lorke and M.~Schleberger,
  \emph{{Nuclear Instruments and Methods in Physics Research Section B: Beam
  Interactions with Materials and Atoms}}, 2016, \textbf{382}, 71--75\relax
\mciteBstWouldAddEndPuncttrue
\mciteSetBstMidEndSepPunct{\mcitedefaultmidpunct}
{\mcitedefaultendpunct}{\mcitedefaultseppunct}\relax
\EndOfBibitem
\bibitem[Kumar \emph{et~al.}(2018)Kumar, Kumar, Tripathi, Tyagi, and
  Avasthi]{Kumar.2018}
S.~Kumar, A.~Kumar, A.~Tripathi, C.~Tyagi and D.~K. Avasthi, \emph{{Journal of
  Applied Physics}}, 2018, \textbf{123}, 161533\relax
\mciteBstWouldAddEndPuncttrue
\mciteSetBstMidEndSepPunct{\mcitedefaultmidpunct}
{\mcitedefaultendpunct}{\mcitedefaultseppunct}\relax
\EndOfBibitem
\bibitem[Ochedowski \emph{et~al.}(2013)Ochedowski, Marinov, Wilbs, Keller,
  Scheuschner, Severin, Bender, Maultzsch, Tegude, and
  Schleberger]{Ochedowski.2013}
O.~Ochedowski, K.~Marinov, G.~Wilbs, G.~Keller, N.~Scheuschner, D.~Severin,
  M.~Bender, J.~Maultzsch, F.~J. Tegude and M.~Schleberger, \emph{{Journal of
  Applied Physics}}, 2013, \textbf{113}, 214306\relax
\mciteBstWouldAddEndPuncttrue
\mciteSetBstMidEndSepPunct{\mcitedefaultmidpunct}
{\mcitedefaultendpunct}{\mcitedefaultseppunct}\relax
\EndOfBibitem
\bibitem[Skopinski \emph{et~al.}(2023)Skopinski, Kretschmer, Ernst, Herder,
  Madau{\ss}, Breuer, Krasheninnikov, and Schleberger]{Skopinski.2023}
L.~Skopinski, S.~Kretschmer, P.~Ernst, M.~Herder, L.~Madau{\ss}, L.~Breuer,
  A.~V. Krasheninnikov and M.~Schleberger, \emph{{Physical Review B}}, 2023,
  \textbf{107}, 044003\relax
\mciteBstWouldAddEndPuncttrue
\mciteSetBstMidEndSepPunct{\mcitedefaultmidpunct}
{\mcitedefaultendpunct}{\mcitedefaultseppunct}\relax
\EndOfBibitem
\bibitem[Lee \emph{et~al.}(2010)Lee, Yan, Brus, Heinz, Hone, and Ryu]{Lee.2010}
C.~Lee, H.~Yan, L.~E. Brus, T.~F. Heinz, J.~Hone and S.~Ryu, \emph{{ACS nano}},
  2010, \textbf{4}, 2695--2700\relax
\mciteBstWouldAddEndPuncttrue
\mciteSetBstMidEndSepPunct{\mcitedefaultmidpunct}
{\mcitedefaultendpunct}{\mcitedefaultseppunct}\relax
\EndOfBibitem
\bibitem[Kim \emph{et~al.}(2017)Kim, Moon, Lee, Choi, Ahmed, Nam, Cho, Shin,
  Park, and Yoo]{Kim.2017}
C.~Kim, I.~Moon, D.~Lee, M.~S. Choi, F.~Ahmed, S.~Nam, Y.~Cho, H.-J. Shin,
  S.~Park and W.~J. Yoo, \emph{{ACS nano}}, 2017, \textbf{11}, 1588--1596\relax
\mciteBstWouldAddEndPuncttrue
\mciteSetBstMidEndSepPunct{\mcitedefaultmidpunct}
{\mcitedefaultendpunct}{\mcitedefaultseppunct}\relax
\EndOfBibitem
\bibitem[Jung \emph{et~al.}(2019)Jung, Choi, Nipane, Borah, Kim, Zangiabadi,
  Taniguchi, Watanabe, Yoo, Hone, and Teherani]{Jung.2019}
Y.~Jung, M.~S. Choi, A.~Nipane, A.~Borah, B.~Kim, A.~Zangiabadi, T.~Taniguchi,
  K.~Watanabe, W.~J. Yoo, J.~Hone and J.~T. Teherani, \emph{{Nature
  Electronics}}, 2019, \textbf{2}, 187--194\relax
\mciteBstWouldAddEndPuncttrue
\mciteSetBstMidEndSepPunct{\mcitedefaultmidpunct}
{\mcitedefaultendpunct}{\mcitedefaultseppunct}\relax
\EndOfBibitem
\bibitem[Guo \emph{et~al.}(2015)Guo, Liu, and Robertson]{Guo.2015}
Y.~Guo, D.~Liu and J.~Robertson, \emph{{ACS Applied Materials {\&}
  Interfaces}}, 2015, \textbf{7}, 25709--25715\relax
\mciteBstWouldAddEndPuncttrue
\mciteSetBstMidEndSepPunct{\mcitedefaultmidpunct}
{\mcitedefaultendpunct}{\mcitedefaultseppunct}\relax
\EndOfBibitem
\bibitem[Smyth \emph{et~al.}(2016)Smyth, Addou, McDonnell, Hinkle, and
  Wallace]{Smyth.2016}
C.~M. Smyth, R.~Addou, S.~McDonnell, C.~L. Hinkle and R.~M. Wallace, \emph{{The
  Journal of Physical Chemistry C}}, 2016, \textbf{120}, 14719--14729\relax
\mciteBstWouldAddEndPuncttrue
\mciteSetBstMidEndSepPunct{\mcitedefaultmidpunct}
{\mcitedefaultendpunct}{\mcitedefaultseppunct}\relax
\EndOfBibitem
\bibitem[Saidi(2014)]{Saidi.2014}
W.~A. Saidi, \emph{{The Journal of Chemical Physics}}, 2014, \textbf{141},
  094707\relax
\mciteBstWouldAddEndPuncttrue
\mciteSetBstMidEndSepPunct{\mcitedefaultmidpunct}
{\mcitedefaultendpunct}{\mcitedefaultseppunct}\relax
\EndOfBibitem
\bibitem[{Di Bartolomeo} \emph{et~al.}(2018){Di Bartolomeo}, Grillo, Urban,
  Iemmo, Giubileo, Luongo, Amato, Croin, Sun, Liang, and
  Ang]{DiBartolomeo.2018b}
A.~{Di Bartolomeo}, A.~Grillo, F.~Urban, L.~Iemmo, F.~Giubileo, G.~Luongo,
  G.~Amato, L.~Croin, L.~Sun, S.-J. Liang and L.~K. Ang, \emph{{Advanced
  Functional Materials}}, 2018, \textbf{28}, 1800657\relax
\mciteBstWouldAddEndPuncttrue
\mciteSetBstMidEndSepPunct{\mcitedefaultmidpunct}
{\mcitedefaultendpunct}{\mcitedefaultseppunct}\relax
\EndOfBibitem
\bibitem[Qiu \emph{et~al.}(2013)Qiu, Xu, Wang, Ren, Nan, Ni, Chen, Yuan, Miao,
  Song, Long, Shi, Sun, Wang, and Wang]{Qiu.2013}
H.~Qiu, T.~Xu, Z.~Wang, W.~Ren, H.~Nan, Z.~Ni, Q.~Chen, S.~Yuan, F.~Miao,
  F.~Song, G.~Long, Y.~Shi, L.~Sun, J.~Wang and X.~Wang, \emph{{Nature
  communications}}, 2013, \textbf{4}, 2642\relax
\mciteBstWouldAddEndPuncttrue
\mciteSetBstMidEndSepPunct{\mcitedefaultmidpunct}
{\mcitedefaultendpunct}{\mcitedefaultseppunct}\relax
\EndOfBibitem
\bibitem[Suh \emph{et~al.}(2014)Suh, Park, Lin, Fu, Park, Jung, Chen, Ko, Jang,
  Sun, Sinclair, Chang, Tongay, and Wu]{Suh.2014}
J.~Suh, T.-E. Park, D.-Y. Lin, D.~Fu, J.~Park, H.~J. Jung, Y.~Chen, C.~Ko,
  C.~Jang, Y.~Sun, R.~Sinclair, J.~Chang, S.~Tongay and J.~Wu, \emph{{Nano
  Letters}}, 2014, \textbf{14}, 6976--6982\relax
\mciteBstWouldAddEndPuncttrue
\mciteSetBstMidEndSepPunct{\mcitedefaultmidpunct}
{\mcitedefaultendpunct}{\mcitedefaultseppunct}\relax
\EndOfBibitem
\bibitem[Tongay \emph{et~al.}(2013)Tongay, Suh, Ataca, Fan, Luce, Kang, Liu,
  Ko, Raghunathanan, Zhou, Ogletree, Li, Grossman, and Wu]{Tongay.2013}
S.~Tongay, J.~Suh, C.~Ataca, W.~Fan, A.~Luce, J.~S. Kang, J.~Liu, C.~Ko,
  R.~Raghunathanan, J.~Zhou, F.~Ogletree, J.~Li, J.~C. Grossman and J.~Wu,
  \emph{{Scientific reports}}, 2013, \textbf{3}, 2657\relax
\mciteBstWouldAddEndPuncttrue
\mciteSetBstMidEndSepPunct{\mcitedefaultmidpunct}
{\mcitedefaultendpunct}{\mcitedefaultseppunct}\relax
\EndOfBibitem
\bibitem[McDonnell \emph{et~al.}(2014)McDonnell, Addou, Buie, Wallace, and
  Hinkle]{McDonnell.2014}
S.~McDonnell, R.~Addou, C.~Buie, R.~M. Wallace and C.~L. Hinkle, \emph{{ACS
  nano}}, 2014, \textbf{8}, 2880--2888\relax
\mciteBstWouldAddEndPuncttrue
\mciteSetBstMidEndSepPunct{\mcitedefaultmidpunct}
{\mcitedefaultendpunct}{\mcitedefaultseppunct}\relax
\EndOfBibitem
\bibitem[Liu \emph{et~al.}(2018)Liu, Guo, Zhu, Liao, Lee, Ding, Shakir, Gambin,
  Huang, and Duan]{Liu.2018}
Y.~Liu, J.~Guo, E.~Zhu, L.~Liao, S.-J. Lee, M.~Ding, I.~Shakir, V.~Gambin,
  Y.~Huang and X.~Duan, \emph{{Nature}}, 2018, \textbf{557}, 696--700\relax
\mciteBstWouldAddEndPuncttrue
\mciteSetBstMidEndSepPunct{\mcitedefaultmidpunct}
{\mcitedefaultendpunct}{\mcitedefaultseppunct}\relax
\EndOfBibitem
\bibitem[Pelella \emph{et~al.}(2020)Pelella, Kharsah, Grillo, Urban,
  Passacantando, Giubileo, Iemmo, Sleziona, Pollmann, Madau{\ss}, Schleberger,
  and {Di Bartolomeo}]{Pelella.2020}
A.~Pelella, O.~Kharsah, A.~Grillo, F.~Urban, M.~Passacantando, F.~Giubileo,
  L.~Iemmo, S.~Sleziona, E.~Pollmann, L.~Madau{\ss}, M.~Schleberger and A.~{Di
  Bartolomeo}, \emph{{ACS Applied Materials {\&} Interfaces}}, 2020,
  \textbf{12}, 40532--40540\relax
\mciteBstWouldAddEndPuncttrue
\mciteSetBstMidEndSepPunct{\mcitedefaultmidpunct}
{\mcitedefaultendpunct}{\mcitedefaultseppunct}\relax
\EndOfBibitem
\bibitem[Shahzad \emph{et~al.}(2019)Shahzad, Jia, Zhao, Wang, Usman, and
  Luo]{Shahzad.2019}
K.~Shahzad, K.~Jia, C.~Zhao, D.~Wang, M.~Usman and J.~Luo, \emph{{Materials}},
  2019, \textbf{12}, 3928\relax
\mciteBstWouldAddEndPuncttrue
\mciteSetBstMidEndSepPunct{\mcitedefaultmidpunct}
{\mcitedefaultendpunct}{\mcitedefaultseppunct}\relax
\EndOfBibitem
\bibitem[Lehtinen \emph{et~al.}(2011)Lehtinen, Kotakoski, Krasheninnikov, and
  Keinonen]{Lehtinen.2011}
O.~Lehtinen, J.~Kotakoski, A.~V. Krasheninnikov and J.~Keinonen,
  \emph{{Nanotechnology}}, 2011, \textbf{22}, 175306\relax
\mciteBstWouldAddEndPuncttrue
\mciteSetBstMidEndSepPunct{\mcitedefaultmidpunct}
{\mcitedefaultendpunct}{\mcitedefaultseppunct}\relax
\EndOfBibitem
\bibitem[Lehtinen \emph{et~al.}(2010)Lehtinen, Kotakoski, Krasheninnikov,
  Tolvanen, Nordlund, and Keinonen]{Lehtinen.2010}
O.~Lehtinen, J.~Kotakoski, A.~V. Krasheninnikov, A.~Tolvanen, K.~Nordlund and
  J.~Keinonen, \emph{{Physical Review B}}, 2010, \textbf{81}, 153401\relax
\mciteBstWouldAddEndPuncttrue
\mciteSetBstMidEndSepPunct{\mcitedefaultmidpunct}
{\mcitedefaultendpunct}{\mcitedefaultseppunct}\relax
\EndOfBibitem
\bibitem[Zhou \emph{et~al.}(2013)Zhou, Zou, Najmaei, Liu, Shi, Kong, Lou,
  Ajayan, Yakobson, and Idrobo]{Zhou.2013}
W.~Zhou, X.~Zou, S.~Najmaei, Z.~Liu, Y.~Shi, J.~Kong, J.~Lou, P.~M. Ajayan,
  B.~I. Yakobson and J.-C. Idrobo, \emph{{Nano Letters}}, 2013, \textbf{13},
  2615--2622\relax
\mciteBstWouldAddEndPuncttrue
\mciteSetBstMidEndSepPunct{\mcitedefaultmidpunct}
{\mcitedefaultendpunct}{\mcitedefaultseppunct}\relax
\EndOfBibitem
\bibitem[Schleberger and Kotakoski(2018)]{Schleberger.2018}
M.~Schleberger and J.~Kotakoski, \emph{{Materials }}, 2018, \textbf{11},
  1885\relax
\mciteBstWouldAddEndPuncttrue
\mciteSetBstMidEndSepPunct{\mcitedefaultmidpunct}
{\mcitedefaultendpunct}{\mcitedefaultseppunct}\relax
\EndOfBibitem
\bibitem[Kozubek \emph{et~al.}(2019)Kozubek, Tripathi, Ghorbani-Asl,
  Kretschmer, Madau{\ss}, Pollmann, O'Brien, McEvoy, Ludacka, Susi, Duesberg,
  Wilhelm, Krasheninnikov, Kotakoski, and Schleberger]{Kozubek.2019}
R.~Kozubek, M.~Tripathi, M.~Ghorbani-Asl, S.~Kretschmer, L.~Madau{\ss},
  E.~Pollmann, M.~O'Brien, N.~McEvoy, U.~Ludacka, T.~Susi, G.~S. Duesberg,
  R.~A. Wilhelm, A.~V. Krasheninnikov, J.~Kotakoski and M.~Schleberger,
  \emph{{The journal of physical chemistry letters}}, 2019, \textbf{10},
  904--910\relax
\mciteBstWouldAddEndPuncttrue
\mciteSetBstMidEndSepPunct{\mcitedefaultmidpunct}
{\mcitedefaultendpunct}{\mcitedefaultseppunct}\relax
\EndOfBibitem
\bibitem[Ghaderzadeh \emph{et~al.}(2020)Ghaderzadeh, Ladygin, Ghorbani-Asl,
  Hlawacek, Schleberger, and Krasheninnikov]{Ghaderzadeh.2020}
S.~Ghaderzadeh, V.~Ladygin, M.~Ghorbani-Asl, G.~Hlawacek, M.~Schleberger and
  A.~V. Krasheninnikov, \emph{{ACS Applied Materials {\&} Interfaces}}, 2020,
  \textbf{12}, 37454--37463\relax
\mciteBstWouldAddEndPuncttrue
\mciteSetBstMidEndSepPunct{\mcitedefaultmidpunct}
{\mcitedefaultendpunct}{\mcitedefaultseppunct}\relax
\EndOfBibitem
\bibitem[Knobloch \emph{et~al.}(2018)Knobloch, Rzepa, Illarionov, Waltl,
  Schanovsky, Stampfer, Furchi, Mueller, and Grasser]{Knobloch.2018}
T.~Knobloch, G.~Rzepa, Y.~Y. Illarionov, M.~Waltl, F.~Schanovsky, B.~Stampfer,
  M.~M. Furchi, T.~Mueller and T.~Grasser, \emph{{IEEE Journal of the Electron
  Devices Society}}, 2018, \textbf{6}, 972--978\relax
\mciteBstWouldAddEndPuncttrue
\mciteSetBstMidEndSepPunct{\mcitedefaultmidpunct}
{\mcitedefaultendpunct}{\mcitedefaultseppunct}\relax
\EndOfBibitem
\bibitem[Nan \emph{et~al.}(2014)Nan, Wang, Wang, Liang, Lu, Chen, He, Tan,
  Miao, Wang, Wang, and Ni]{Nan.2014}
H.~Nan, Z.~Wang, W.~Wang, Z.~Liang, Y.~Lu, Q.~Chen, D.~He, P.~Tan, F.~Miao,
  X.~Wang, J.~Wang and Z.~Ni, \emph{{ACS nano}}, 2014, \textbf{8},
  5738--5745\relax
\mciteBstWouldAddEndPuncttrue
\mciteSetBstMidEndSepPunct{\mcitedefaultmidpunct}
{\mcitedefaultendpunct}{\mcitedefaultseppunct}\relax
\EndOfBibitem
\bibitem[Tongay \emph{et~al.}(2013)Tongay, Zhou, Ataca, Liu, Kang, Matthews,
  You, Li, Grossman, and Wu]{Tongay.2013b}
S.~Tongay, J.~Zhou, C.~Ataca, J.~Liu, J.~S. Kang, T.~S. Matthews, L.~You,
  J.~Li, J.~C. Grossman and J.~Wu, \emph{{Nano Letters}}, 2013, \textbf{13},
  2831--2836\relax
\mciteBstWouldAddEndPuncttrue
\mciteSetBstMidEndSepPunct{\mcitedefaultmidpunct}
{\mcitedefaultendpunct}{\mcitedefaultseppunct}\relax
\EndOfBibitem
\bibitem[Wang \emph{et~al.}(2022)Wang, Zeng, Zhou, Lu, Hu, Wang, Wang, Xiao,
  Wang, Chen, Xu, Zhang, and Bao]{Wang.2022}
S.~Wang, X.~Zeng, Y.~Zhou, J.~Lu, Y.~Hu, W.~Wang, J.~Wang, Y.~Xiao, X.~Wang,
  D.~Chen, T.~Xu, M.~Zhang and X.~Bao, \emph{{ACS Applied Electronic
  Materials}}, 2022, \textbf{4}, 955--963\relax
\mciteBstWouldAddEndPuncttrue
\mciteSetBstMidEndSepPunct{\mcitedefaultmidpunct}
{\mcitedefaultendpunct}{\mcitedefaultseppunct}\relax
\EndOfBibitem
\bibitem[Wei \emph{et~al.}(2014)Wei, Yu, Hu, Cheng, Yu, Wang, Xiao, Wang, Wang,
  and Shi]{Wei.2014}
X.~Wei, Z.~Yu, F.~Hu, Y.~Cheng, L.~Yu, X.~Wang, M.~Xiao, J.~Wang, X.~Wang and
  Y.~Shi, \emph{{AIP Advances}}, 2014, \textbf{4}, 123004\relax
\mciteBstWouldAddEndPuncttrue
\mciteSetBstMidEndSepPunct{\mcitedefaultmidpunct}
{\mcitedefaultendpunct}{\mcitedefaultseppunct}\relax
\EndOfBibitem
\bibitem[Kong \emph{et~al.}(2014)Kong, Enders, Rahman, and Dowben]{Kong.2014}
L.~Kong, A.~Enders, T.~S. Rahman and P.~A. Dowben, \emph{{Journal of physics.
  Condensed matter : An Institute of Physics journal}}, 2014, \textbf{26},
  443001\relax
\mciteBstWouldAddEndPuncttrue
\mciteSetBstMidEndSepPunct{\mcitedefaultmidpunct}
{\mcitedefaultendpunct}{\mcitedefaultseppunct}\relax
\EndOfBibitem
\bibitem[KC \emph{et~al.}(2015)KC, Longo, Wallace, and Cho]{KC.2015}
S.~KC, R.~C. Longo, R.~M. Wallace and K.~Cho, \emph{{Journal of Applied
  Physics}}, 2015, \textbf{117}, 135301\relax
\mciteBstWouldAddEndPuncttrue
\mciteSetBstMidEndSepPunct{\mcitedefaultmidpunct}
{\mcitedefaultendpunct}{\mcitedefaultseppunct}\relax
\EndOfBibitem
\bibitem[Degraeve \emph{et~al.}(2008)Degraeve, Cho, Govoreanu, Kaczer, Zahid,
  {van Houdt}, Jurczak, and Groeseneken]{Degraeve.2008}
R.~Degraeve, M.~Cho, B.~Govoreanu, B.~Kaczer, M.~B. Zahid, J.~{van Houdt},
  M.~Jurczak and G.~Groeseneken, {IEEE International Electron Devices Meeting,
  2008}, Piscataway, NJ, 2008, pp. 1--4\relax
\mciteBstWouldAddEndPuncttrue
\mciteSetBstMidEndSepPunct{\mcitedefaultmidpunct}
{\mcitedefaultendpunct}{\mcitedefaultseppunct}\relax
\EndOfBibitem
\bibitem[Illarionov \emph{et~al.}(2017)Illarionov, Knobloch, Waltl, Rzepa,
  Pospischil, Polyushkin, Furchi, Mueller, and Grasser]{Illarionov.2017}
Y.~Y. Illarionov, T.~Knobloch, M.~Waltl, G.~Rzepa, A.~Pospischil, D.~K.
  Polyushkin, M.~M. Furchi, T.~Mueller and T.~Grasser, \emph{{2D Materials}},
  2017, \textbf{4}, 025108\relax
\mciteBstWouldAddEndPuncttrue
\mciteSetBstMidEndSepPunct{\mcitedefaultmidpunct}
{\mcitedefaultendpunct}{\mcitedefaultseppunct}\relax
\EndOfBibitem
\bibitem[Huang \emph{et~al.}(2020)Huang, Zhao, Sun, Liu, Liu, Chang, Zeng, and
  Liu]{Huang.2020}
K.~Huang, M.~Zhao, B.~Sun, X.~Liu, J.~Liu, H.~Chang, Y.~Zeng and H.~Liu,
  \emph{{ACS Applied Electronic Materials}}, 2020, \textbf{2}, 971--979\relax
\mciteBstWouldAddEndPuncttrue
\mciteSetBstMidEndSepPunct{\mcitedefaultmidpunct}
{\mcitedefaultendpunct}{\mcitedefaultseppunct}\relax
\EndOfBibitem
\bibitem[Novoselov \emph{et~al.}(2005)Novoselov, Jiang, Schedin, Booth,
  Khotkevich, Morozov, and Geim]{Novoselov.2005}
K.~S. Novoselov, D.~Jiang, F.~Schedin, T.~J. Booth, V.~V. Khotkevich, S.~V.
  Morozov and A.~K. Geim, \emph{{Proceedings of the National Academy of
  Sciences of the United States of America}}, 2005, \textbf{102},
  10451--10453\relax
\mciteBstWouldAddEndPuncttrue
\mciteSetBstMidEndSepPunct{\mcitedefaultmidpunct}
{\mcitedefaultendpunct}{\mcitedefaultseppunct}\relax
\EndOfBibitem
\bibitem[Amani \emph{et~al.}(2013)Amani, Chin, Birdwell, O'Regan, Najmaei, Liu,
  Ajayan, Lou, and Dubey]{Amani.2013}
M.~Amani, M.~L. Chin, A.~G. Birdwell, T.~P. O'Regan, S.~Najmaei, Z.~Liu, P.~M.
  Ajayan, J.~Lou and M.~Dubey, \emph{{Applied Physics Letters}}, 2013,
  \textbf{102}, 193107\relax
\mciteBstWouldAddEndPuncttrue
\mciteSetBstMidEndSepPunct{\mcitedefaultmidpunct}
{\mcitedefaultendpunct}{\mcitedefaultseppunct}\relax
\EndOfBibitem
\bibitem[Chakraborty \emph{et~al.}(2012)Chakraborty, Bera, Muthu, Bhowmick,
  Waghmare, and Sood]{Chakraborty.2012}
B.~Chakraborty, A.~Bera, D.~V.~S. Muthu, S.~Bhowmick, U.~V. Waghmare and A.~K.
  Sood, \emph{{Physical Review B}}, 2012, \textbf{85}, 161403\relax
\mciteBstWouldAddEndPuncttrue
\mciteSetBstMidEndSepPunct{\mcitedefaultmidpunct}
{\mcitedefaultendpunct}{\mcitedefaultseppunct}\relax
\EndOfBibitem
\bibitem[Pollmann \emph{et~al.}(2020)Pollmann, Madau{\ss}, Schumacher, Kumar,
  Heuvel, {vom Ende}, Yilmaz, G{\"u}ng{\"o}rm{\"u}s, and
  Schleberger]{Pollmann.2020}
E.~Pollmann, L.~Madau{\ss}, S.~Schumacher, U.~Kumar, F.~Heuvel, C.~{vom Ende},
  S.~Yilmaz, S.~G{\"u}ng{\"o}rm{\"u}s and M.~Schleberger,
  \emph{{Nanotechnology}}, 2020, \textbf{31}, 505604\relax
\mciteBstWouldAddEndPuncttrue
\mciteSetBstMidEndSepPunct{\mcitedefaultmidpunct}
{\mcitedefaultendpunct}{\mcitedefaultseppunct}\relax
\EndOfBibitem
\bibitem[Guo \emph{et~al.}(2015)Guo, Wei, Shu, Liu, Yin, Guan, Han, Gao, and
  Chen]{Guo.2015b}
Y.~Guo, X.~Wei, J.~Shu, B.~Liu, J.~Yin, C.~Guan, Y.~Han, S.~Gao and Q.~Chen,
  \emph{{Applied Physics Letters}}, 2015, \textbf{106}, 103109\relax
\mciteBstWouldAddEndPuncttrue
\mciteSetBstMidEndSepPunct{\mcitedefaultmidpunct}
{\mcitedefaultendpunct}{\mcitedefaultseppunct}\relax
\EndOfBibitem
\bibitem[Choi \emph{et~al.}(2015)Choi, Raza, Lee, Jeon, Pezeshki, Min, Kim,
  Yoon, Ju, Lee, and Im]{Choi.2015}
K.~Choi, S.~R.~A. Raza, H.~S. Lee, P.~J. Jeon, A.~Pezeshki, S.-W. Min, J.~S.
  Kim, W.~Yoon, S.-Y. Ju, K.~Lee and S.~Im, \emph{{Nanoscale}}, 2015,
  \textbf{7}, 5617--5623\relax
\mciteBstWouldAddEndPuncttrue
\mciteSetBstMidEndSepPunct{\mcitedefaultmidpunct}
{\mcitedefaultendpunct}{\mcitedefaultseppunct}\relax
\EndOfBibitem
\bibitem[Illarionov \emph{et~al.}(2016)Illarionov, Rzepa, Waltl, Knobloch,
  Grill, Furchi, Mueller, and Grasser]{Illarionov.2016}
Y.~Y. Illarionov, G.~Rzepa, M.~Waltl, T.~Knobloch, A.~Grill, M.~M. Furchi,
  T.~Mueller and T.~Grasser, \emph{{2D Materials}}, 2016, \textbf{3},
  035004\relax
\mciteBstWouldAddEndPuncttrue
\mciteSetBstMidEndSepPunct{\mcitedefaultmidpunct}
{\mcitedefaultendpunct}{\mcitedefaultseppunct}\relax
\EndOfBibitem
\bibitem[Zhang \emph{et~al.}(2015)Zhang, Wang, Zhang, Jin, Zhu, Sun, Zhang,
  Zhou, and Xiu]{Zhang.2015}
E.~Zhang, W.~Wang, C.~Zhang, Y.~Jin, G.~Zhu, Q.~Sun, D.~W. Zhang, P.~Zhou and
  F.~Xiu, \emph{{ACS nano}}, 2015, \textbf{9}, 612--619\relax
\mciteBstWouldAddEndPuncttrue
\mciteSetBstMidEndSepPunct{\mcitedefaultmidpunct}
{\mcitedefaultendpunct}{\mcitedefaultseppunct}\relax
\EndOfBibitem
\bibitem[Hou \emph{et~al.}(2018)Hou, Yan, Liu, Ding, Zhang, and Zhou]{Hou.2018}
X.~Hou, X.~Yan, C.~Liu, S.~Ding, D.~W. Zhang and P.~Zhou, \emph{{Semiconductor
  Science and Technology}}, 2018, \textbf{33}, 034001\relax
\mciteBstWouldAddEndPuncttrue
\mciteSetBstMidEndSepPunct{\mcitedefaultmidpunct}
{\mcitedefaultendpunct}{\mcitedefaultseppunct}\relax
\EndOfBibitem
\bibitem[Nagumo \emph{et~al.}(06.12.2010 - 08.12.2010)Nagumo, Takeuchi, Hase,
  and Hayashi]{Nagumo.06.12.201008.12.2010}
T.~Nagumo, K.~Takeuchi, T.~Hase and Y.~Hayashi, {2010 International Electron
  Devices Meeting}, 06.12.2010 - 08.12.2010, pp. 28.3.1--28.3.4\relax
\mciteBstWouldAddEndPuncttrue
\mciteSetBstMidEndSepPunct{\mcitedefaultmidpunct}
{\mcitedefaultendpunct}{\mcitedefaultseppunct}\relax
\EndOfBibitem
\bibitem[Lee \emph{et~al.}(2011)Lee, Kang, Jung, Kim, Hwang, Chung, Seo, Choi,
  and Lee]{Lee.2011}
Y.~G. Lee, C.~G. Kang, U.~J. Jung, J.~J. Kim, H.~J. Hwang, H.-J. Chung, S.~Seo,
  R.~Choi and B.~H. Lee, \emph{{Applied Physics Letters}}, 2011, \textbf{98},
  183508\relax
\mciteBstWouldAddEndPuncttrue
\mciteSetBstMidEndSepPunct{\mcitedefaultmidpunct}
{\mcitedefaultendpunct}{\mcitedefaultseppunct}\relax
\EndOfBibitem
\bibitem[Kumar \emph{et~al.}(2023)Kumar, Viscardi, Faella, Giubileo, Intonti,
  Pelella, Sleziona, Kharsah, Schleberger, and {Di Bartolomeo}]{Kumar.2023}
A.~Kumar, L.~Viscardi, E.~Faella, F.~Giubileo, K.~Intonti, A.~Pelella,
  S.~Sleziona, O.~Kharsah, M.~Schleberger and A.~{Di Bartolomeo},
  \emph{{Journal of Materials Science}}, 2023, \textbf{58}, 2689--2699\relax
\mciteBstWouldAddEndPuncttrue
\mciteSetBstMidEndSepPunct{\mcitedefaultmidpunct}
{\mcitedefaultendpunct}{\mcitedefaultseppunct}\relax
\EndOfBibitem
\bibitem[Ziegler \emph{et~al.}(2010)Ziegler, Ziegler, and
  Biersack]{Ziegler.2010}
J.~F. Ziegler, M.~D. Ziegler and J.~P. Biersack, \emph{{Nuclear Instruments and
  Methods in Physics Research Section B: Beam Interactions with Materials and
  Atoms}}, 2010, \textbf{268}, 1818--1823\relax
\mciteBstWouldAddEndPuncttrue
\mciteSetBstMidEndSepPunct{\mcitedefaultmidpunct}
{\mcitedefaultendpunct}{\mcitedefaultseppunct}\relax
\EndOfBibitem
\bibitem[Zschornack \emph{et~al.}(2008)Zschornack, Kreller, Ovsyannikov,
  Grossman, Kentsch, Schmidt, Ullmann, and Heller]{Zschornack.2008}
G.~Zschornack, M.~Kreller, V.~P. Ovsyannikov, F.~Grossman, U.~Kentsch,
  M.~Schmidt, F.~Ullmann and R.~Heller, \emph{{The Review of scientific
  instruments}}, 2008, \textbf{79}, 02A703\relax
\mciteBstWouldAddEndPuncttrue
\mciteSetBstMidEndSepPunct{\mcitedefaultmidpunct}
{\mcitedefaultendpunct}{\mcitedefaultseppunct}\relax
\EndOfBibitem
\end{mcitethebibliography}
\bibliographystyle{rsc} 

\end{document}